\documentclass[prd,twocolumn,showpacs,groupedaddress,superscriptaddress,amsmath,amssymb]{revtex4-2} 
 
\usepackage[utf8]{inputenc}
\usepackage[english]{babel}	

\usepackage{stfloats}

\usepackage{cmap}
\usepackage{bbm}
\usepackage{fontenc}

\usepackage{enumerate}
\usepackage{multirow}
\usepackage{float}
\usepackage{comment}
\usepackage{lineno}

\usepackage{amsmath,amsfonts,amssymb,mathtools}	
\usepackage{euscript} 
\usepackage{mathrsfs}
\usepackage{mathtext}

\usepackage{color}			
\usepackage{graphicx}
\usepackage[export]{adjustbox}
\DeclareGraphicsExtensions{.png,.jpg}
\graphicspath{}	
\usepackage[section]{placeins} 

\usepackage{physics}
\usepackage{braket}
\usepackage{xparse}	

\usepackage{slashed}
\usepackage{indentfirst} 
\usepackage{enumitem}

\usepackage[colorlinks=true, filecolor=blue, citecolor=blue,urlcolor=black]{hyperref}
\usepackage{orcidlink}
\usepackage{url}
\usepackage{cleveref}
\date{\today}

\begin{document}

\title{Shedding light on Dark Sectors with high-energy muons at the NA64 experiment at the CERN SPS} 

\author{Yu.~M.~Andreev\orcidlink{0000-0002-7397-9665}}
\affiliation{Authors affiliated with an institute covered by a cooperation agreement with CERN}
\author{D.~Banerjee\orcidlink{0000-0003-0531-1679}}
\affiliation{CERN, European Organization for Nuclear Research, CH-1211 Geneva, Switzerland}
\author{B.~Banto Oberhauser\orcidlink{0009-0006-4795-1008}}
\affiliation{ETH Z\"urich, Institute for Particle Physics and Astrophysics, CH-8093 Z\"urich, Switzerland}
\author{J.~Bernhard\orcidlink{0000-0001-9256-971X}}
\affiliation{CERN, European Organization for Nuclear Research, CH-1211 Geneva, Switzerland}
\author{P.~Bisio\orcidlink{/0009-0006-8677-7495}}
\affiliation{INFN, Sezione di Genova, 16147 Genova, Italia}
\affiliation{Universit\`a degli Studi di Genova, 16126 Genova, Italia}
\author{N.~Charitonidis\orcidlink{0000-0001-9506-1022}}
\affiliation{CERN, European Organization for Nuclear Research, CH-1211 Geneva, Switzerland}
\author{P.~Crivelli\orcidlink{0000-0001-5430-9394}}
\affiliation{ETH Z\"urich, Institute for Particle Physics and Astrophysics, CH-8093 Z\"urich, Switzerland}
\author{E.~Depero\orcidlink{0000-0003-2239-1746}}
\affiliation{ETH Z\"urich, Institute for Particle Physics and Astrophysics, CH-8093 Z\"urich, Switzerland}
\author{A.~V.~Dermenev\orcidlink{0000-0001-5619-376X}}
\affiliation{Authors affiliated with an institute covered by a cooperation agreement with CERN}
\author{S.~V.~Donskov\orcidlink{0000-0002-3988-7687}}
\affiliation{Authors affiliated with an institute covered by a cooperation agreement with CERN}
\author{R.~R.~Dusaev\orcidlink{0000-0002-6147-8038}}
\affiliation{Authors affiliated with an institute covered by a cooperation agreement with CERN}
\author{T.~Enik\orcidlink{0000-0002-2761-9730}}
\affiliation{Authors affiliated with an international laboratory covered by a cooperation agreement with CERN}
\author{V.~N.~Frolov}
\affiliation{Authors affiliated with an international laboratory covered by a cooperation agreement with CERN}
\author{A.~Gardikiotis\orcidlink{0000-0002-4435-2695}}
\affiliation{Physics Department, University of Patras, 265 04 Patras, Greece}
\author{S.~V.~Gertsenberger\orcidlink{0009-0006-1640-9443}}
\affiliation{Authors affiliated with an international laboratory covered by a cooperation agreement with CERN}
\author{S. Girod}
\affiliation{CERN, European Organization for Nuclear Research, CH-1211 Geneva, Switzerland}
\author{S.~N.~Gninenko\orcidlink{0000-0001-6495-7619}}
\affiliation{Authors affiliated with an institute covered by a cooperation agreement with CERN}
\affiliation{Center for Theoretical and Experimental Particle Physics, Facultad de Ciencias Exactas, Universidad Andres Bello, Fernandez Concha 700, Santiago, Chile}
\author{M.~H\"osgen}
\affiliation{Universit\"at Bonn, Helmholtz-Institut f\"ur Strahlen-und Kernphysik, 53115 Bonn, Germany}
\author{R. Joosten}
\affiliation{Rheinische Friedrich-Wilhelms-Universität, Bonn, Germany}
\author{V.~A.~Kachanov\orcidlink{0000-0002-3062-010X}}
\affiliation{Authors affiliated with an institute covered by a cooperation agreement with CERN}
\author{Y.~Kambar\orcidlink{0009-0000-9185-2353}}
\affiliation{Authors affiliated with an international laboratory covered by a cooperation agreement with CERN}
\author{A.~E.~Karneyeu\orcidlink{0000-0001-9983-1004}}
\affiliation{Authors affiliated with an institute covered by a cooperation agreement with CERN}
\author{E.~A.~Kasianova}
\affiliation{Authors affiliated with an international laboratory covered by a cooperation agreement with CERN}
\author{G.~Kekelidze\orcidlink{0000-0002-5393-9199}}
\affiliation{Authors affiliated with an international laboratory covered by a cooperation agreement with CERN}
\author{B.~Ketzer\orcidlink{0000-0002-3493-3891}}
\affiliation{Universit\"at Bonn, Helmholtz-Institut f\"ur Strahlen-und Kernphysik, 53115 Bonn, Germany}
\author{D.~V.~Kirpichnikov\orcidlink{0000-0002-7177-077X}}
\affiliation{Authors affiliated with an institute covered by a cooperation agreement with CERN}
\author{M.~M.~Kirsanov\orcidlink{0000-0002-8879-6538}}
\affiliation{Authors affiliated with an institute covered by a cooperation agreement with CERN}
\author{V.~N.~Kolosov}
\affiliation{Authors affiliated with an institute covered by a cooperation agreement with CERN}
\author{V.~A.~Kramarenko\orcidlink{0000-0002-8625-5586}}
\affiliation{Authors affiliated with an institute covered by a cooperation agreement with CERN}
\affiliation{Authors affiliated with an international laboratory covered by a cooperation agreement with CERN}
\author{L.~V.~Kravchuk\orcidlink{0000-0001-8631-4200}}
\affiliation{Authors affiliated with an institute covered by a cooperation agreement with CERN}
\author{N.~V.~Krasnikov\orcidlink{0000-0002-8717-6492}}
\affiliation{Authors affiliated with an institute covered by a cooperation agreement with CERN}
\affiliation{Authors affiliated with an international laboratory covered by a cooperation agreement with CERN}
\author{S.~V.~Kuleshov\orcidlink{0000-0002-3065-326X}}
\affiliation{Center for Theoretical and Experimental Particle Physics, Facultad de Ciencias Exactas, Universidad Andres Bello, Fernandez Concha 700, Santiago, Chile}
\affiliation{Millennium Institute for Subatomic Physics at High-Energy Frontier (SAPHIR), Fernandez Concha 700, Santiago, Chile}
\author{V.~E.~Lyubovitskij\orcidlink{0000-0001-7467-572X}}
\affiliation{Authors affiliated with an institute covered by a cooperation agreement with CERN}
\affiliation{Universidad T\'ecnica Federico Santa Mar\'ia and CCTVal, 2390123 Valpara\'iso, Chile}
\affiliation{Millennium Institute for Subatomic Physics at High-Energy Frontier (SAPHIR), Fernandez Concha 700, Santiago, Chile}
\author{V.~Lysan\orcidlink{0009-0004-1795-1651}}
\affiliation{Authors affiliated with an international laboratory covered by a cooperation agreement with CERN}
\author{V.~A.~Matveev\orcidlink{0000-0002-2745-5908}}
\affiliation{Authors affiliated with an international laboratory covered by a cooperation agreement with CERN}
\author{R.~Mena~Fredes}
\affiliation{Millennium Institute for Subatomic Physics at High-Energy Frontier (SAPHIR), Fernandez Concha 700, Santiago, Chile}
\affiliation{Universidad T\'ecnica Federico Santa Mar\'ia and CCTVal, 2390123 Valpara\'iso, Chile}
\author{R.~ G.~Mena~Yanssen}
\affiliation{Millennium Institute for Subatomic Physics at High-Energy Frontier (SAPHIR), Fernandez Concha 700, Santiago, Chile}
\affiliation{Universidad T\'ecnica Federico Santa Mar\'ia and CCTVal, 2390123 Valpara\'iso, Chile}
\author{L.~Molina Bueno\orcidlink{0000-0001-9720-9764}}
\affiliation{Instituto de Fisica Corpuscular (CSIC/UV), Carrer del Catedratic Jose Beltran Martinez, 2, 46980 Paterna, Valencia, Spain}
\author{M.~Mongillo\orcidlink{0009-0000-7331-4076}}
\affiliation{ETH Z\"urich, Institute for Particle Physics and Astrophysics, CH-8093 Z\"urich, Switzerland}
\author{D.~V.~Peshekhonov\orcidlink{0009-0008-9018-5884}}
\affiliation{Authors affiliated with an international laboratory covered by a cooperation agreement with CERN}
\author{V.~A.~Polyakov\orcidlink{0000-0001-5989-0990}}
\affiliation{Authors affiliated with an institute covered by a cooperation agreement with CERN}
\author{B.~Radics\orcidlink{0000-0002-8978-1725}}
\affiliation{York University, Toronto, Canada}
\author{K.~M.~Salamatin\orcidlink{0000-0001-6287-8685}}
\affiliation{Authors affiliated with an international laboratory covered by a cooperation agreement with CERN}
\author{V.~D.~Samoylenko}
\affiliation{Authors affiliated with an institute covered by a cooperation agreement with CERN}
\author{D.~A.~Shchukin\orcidlink{0009-0007-5508-3615}}
\affiliation{Authors affiliated with an institute covered by a cooperation agreement with CERN}
\author{O.~Soto}
\affiliation{Departamento de Fisica, Facultad de Ciencias, Universidad de La Serena, Avenida Cisternas 1200, La Serena, Chile}
\affiliation{Millennium Institute for Subatomic Physics at High-Energy Frontier (SAPHIR), Fernandez Concha 700, Santiago, Chile}
\author{H.~Sieber\orcidlink{0000-0003-1476-4258}}
\email[\textbf{Corresponding author}:]{henri.hugo.sieber@cern.ch}
\affiliation{ETH Z\"urich, Institute for Particle Physics and Astrophysics, CH-8093 Z\"urich, Switzerland}
\author{V.~O.~Tikhomirov\orcidlink{0000-0002-9634-0581}}
\affiliation{Authors affiliated with an institute covered by a cooperation agreement with CERN}
\author{I.~V.~Tlisova\orcidlink{0000-0003-1552-2015}}
\affiliation{Authors affiliated with an institute covered by a cooperation agreement with CERN}
\author{A.~N.~Toropin\orcidlink{0000-0002-2106-4041}}
\affiliation{Authors affiliated with an institute covered by a cooperation agreement with CERN}
\author{M.~Tuzi\orcidlink{0009-0000-6276-1401}}
\affiliation{Instituto de Fisica Corpuscular (CSIC/UV), Carrer del Catedratic Jose Beltran Martinez, 2, 46980 Paterna, Valencia, Spain}
\author{B. M. Veit}
\affiliation{Johannes Gutenberg Universitaet Mainz, Germany}
\author{P.~V.~Volkov\orcidlink{0000-0002-7668-3691}}
\affiliation{Authors affiliated with an international laboratory covered by a cooperation agreement with CERN}
\author{V.~Yu.~Volkov\orcidlink{0009-0005-3500-5121}}
\affiliation{Authors affiliated with an institute covered by a cooperation agreement with CERN}
\author{I.~V.~Voronchikhin\orcidlink{0000-0003-3037-636X}}
\affiliation{Authors affiliated with an institute covered by a cooperation agreement with CERN}
\author{J.~Zamora-Sa\'a\orcidlink{0000-0002-5030-7516}}
\affiliation{Center for Theoretical and Experimental Particle Physics, Facultad de Ciencias Exactas, Universidad Andres Bello, Fernandez Concha 700, Santiago, Chile}
\affiliation{Millennium Institute for Subatomic Physics at High-Energy Frontier (SAPHIR), Fernandez Concha 700, Santiago, Chile}
\author{A.~S.~Zhevlakov\orcidlink{0000-0002-7775-5917}}
\affiliation{Authors affiliated with an international laboratory covered by a cooperation agreement with CERN}

\begin{abstract}
A search for Dark Sectors is performed using the unique M2 beam line at the CERN Super Proton Synchrotron. New particles ($X$) could be produced in the bremsstrahlung-like reaction of high energy 160 GeV muons impinging on an active target, $\mu N\rightarrow\mu NX$, followed by their decays, $X\rightarrow\text{invisible}$. The experimental signature would be a scattered single muon from the target, with about less than half of its initial energy and no activity in the sub-detectors located downstream the interaction point. The full sample of the 2022 run is analyzed through the missing energy/momentum channel, with a total statistics of $(1.98\pm0.02)\times10^{10}$ muons on target. We demonstrate that various muon-philic scenarios involving different types of mediators, such as scalar or vector particles, can be probed simultaneously with such a technique. For the vector-case, besides a $L_\mu-L_\tau$ $Z'$ vector boson, we also consider an invisibly decaying dark photon ($A'\rightarrow\text{invisible}$). This search is complementary to NA64 running with electrons and positrons, thus, opening the possibility to expand the exploration of the thermal light dark matter parameter space by combining the results obtained with the three beams.
\end{abstract}
\maketitle
\section{Introduction}\label{sec:introduction}
The observational pieces of evidence for the existence of Dark Matter (DM) are derived from its gravitational nature, including galaxy rotational curves \cite{Rubin:1970zza,Freeman:1970mx,1985ApJ...295..305V,Begeman:1991iy}, gravitational lensing \cite{Zwicky:1937zza,Tyson:1998vp,Clowe:2006eq}, Cosmic Microwave Background (CMB) anisotropies \cite{Planck:2013pxb,Planck:2018vyg}, or large-scale structure (LSS) formation \cite{Refregier:2003ct,2dFGRS:2005yhx,DES:2021wwk}. Together, they suggest that the relic density of DM, denoted $\Omega_\text{DM}\simeq0.27$, surpasses that of baryonic matter by a factor of approximately six. \\ \indent
However, fundamental aspects concerning the intrinsic nature of DM and the origin of its relic abundance persist as open questions. The apparent dearth of significant interactions with Standard Model (SM) particles could be explained in the framework of Dark Sectors (DS), with a particle and field content, which is singlet under the SM gauge group. In this scenario, DM is a part of the DS and is charged under a new force responsible for the interaction with SM particles through a corresponding mediator \cite{Kobzarev:1966qya,Blinnikov:1982eh,Foot:1991bp,Hodges:1993yb,Berezhiani:1995am}. While the canonical DS model is associated with a new light gauge vector boson from a broken $U(1)_D$ symmetry \cite{Holdom:1985ag}, the dark photon $A'$, new portal scenarios with gauged SM symmetries have gained popular interest. Those postulate in particular new forces mediated by gauge bosons,  $Z'$, associated with gauged symmetries such as $U(1)_{B-L}$, $U(1)_{B-3L}$, $U(1)_{L_m-L_n}$ or $U(1)_B$ \cite{Davidson:1978pm,Marshak:1979fm,He:1990pn,He:1991qd,Foot:1994vd} with respectively $B$ the baryon number and $L_m$, $m=e,\mu,\tau$, the lepton number for the different generations. Within those models, the mediator $Z'$ is assumed to couple to DM with coupling $g_\chi\sim\mathcal{O}(1)$, and to SM particles through $g=\epsilon e$, with $e$ the electric charge and $\epsilon$ the coupling strength, $\epsilon\ll 1$. Other well-motivated scenarios involve in particular the exchange between DM and SM particles of lepto-philic spin-0 mediators with either scalar or pseudo-scalar interactions, or the existence of millicharged particles (see e.g. \cite{Berlin:2018bsc} for a complete review of New Physics (NP) scenarios associated with DM). \\ \indent
An attractive framework belongs to the class of models with anomaly-free lepton numbers used in gauge symmetries, such as $L_\mu-L_\tau$, corresponding to the SM gauge group extension $SU(3)_c\bigotimes SU(2)_{L}\bigotimes U(1)_{Y}\bigotimes U(1)_{L_\mu-L_\tau}$. As such, the $Z'$ vector boson from the broken $U(1)_{L_\mu-L_\tau}$ symmetry couples directly to the second and third lepton generations, and their corresponding left-handed neutrinos \cite{He:1990pn,He:1991qd,Foot:1994vd,Altmannshofer:2016jzy,Kile:2014jea,Park:2015gdo}. Our framework is based on the following Lagrangian,
\begin{equation}
\label{eq:lagrangian}
\mathcal{L}\supset-\frac{1}{4}F_{\alpha\beta}^\prime F^{\alpha\beta\prime}+\frac{m_{Z^\prime}^2}{2}Z_\alpha^\prime Z^{\alpha\prime}-g_{Z^\prime}Z_\alpha^{\prime}J_{\mu-\tau}^\alpha,
\end{equation}
where $J_{\mu-\tau}^\alpha$ is the $U(1)_{L_\mu-L_\tau}$ leptonic current, 
\begin{equation}
    J_{\mu-\tau}^\alpha=\big(\bar{\mu}\gamma^\alpha\mu-\bar{\tau}\gamma^\alpha\tau
        +\bar{\nu}_\mu\gamma^\alpha P_L\nu_\mu
        -\bar{\nu}_\tau\gamma^\alpha P_L\nu_\tau\big),
\end{equation}
and $F_{\alpha\beta}^\prime$ the field strength tensor associated with the massive vector field $Z_\alpha^{\prime}$, $g_{Z'}=\epsilon_{Z'}e$ the coupling of $Z'$ to SM particles, and $P_L$ the left-handed chiral projection operator. Within this \emph{vanilla} model, the gauge boson $Z'$ decays invisibly to SM neutrinos, such that 
\begin{equation}
    \label{eq:decays-neutrinos}
    \Gamma(Z'\rightarrow\overline{\nu}{\nu})=\frac{\alpha_{Z'}m_{Z'}}{3},
\end{equation}
with $\alpha_{Z'}=g_{Z'}^2/(4\pi)$. At $m_{Z'}>2m_\mu$ the visible decays to SM leptons, $Z'\rightarrow\bar{\mu}\mu$, open.\\ \indent
The extension of Eq. \eqref{eq:lagrangian} to interactions with DM candidates, being consistent in predicting their thermal history \cite{Feng:2008mu,Feng:2008ya} through $\langle\sigma v\rangle\simeq1\ \text{pb}\times c\simeq(1-3)\times10^{-26}$ cm$^3$s$^{-1}$ \cite{Arcadi:2017kky,Planck:2018vyg}, is achieved by adding a term of the type $\mathcal{L}\supseteq -g_\chi Z_\alpha^\prime J_\chi^\alpha$ with $J_\chi^\alpha$ a DS current reading
\begin{equation}
    J_{\chi}^{\alpha}=g_{\chi}
    \begin{cases}
        i\chi^{\ast}\partial^\alpha\chi+\text{h.c.}, & \text{complex scalar} \\
        1/2\overline{\chi}\gamma^\alpha\gamma^{5}\chi, & \text{Majorana}\\
        i\overline{\chi}_1\gamma^\alpha\chi_2, & \text{pseudo-Dirac} \\
        \overline{\chi}\gamma^\alpha\chi, & \text{Dirac}\\
    \end{cases},
\end{equation}
and the coupling $g_\chi$ to the DM candidates. In the case where $m_{Z^\prime}>m_\chi$ (away from the near on-shell resonant enhancement $m_{Z^\prime}\simeq2m_{\chi}$), the relic density is set by $\bar{\chi}\chi(\rightarrow Z^{(\ast)\prime}\rightarrow)\bar{f}f$, $f=\mu,\tau,\nu$, with the relevant $s-$channel annihilation cross-section scaling as \cite{Altmannshofer:2016jzy,Arcadi:2017kky,Kirpichnikov:2020tcf} $\langle\sigma v\rangle\propto(g_\chi g_{Z'})^2m_\chi^2/m_{Z^\prime}^4=y m_\chi^{-2}$ for Dirac DM. This defines the dimensionless $y$ parameter for probing the DM relic abundance such that
\begin{equation}
    \label{eq:dimensionless-y}
    y=(g_\chi g_{Z'})^2\left(\frac{m_\chi}{m_{Z'}}\right)^4.    
\end{equation}
In particular, in the case where $m_{Z'}>2m_\chi$, the light vector boson predominantly decays into DM candidates, provided that the coupling $g_\chi>g_{Z'}$, and the corresponding decay width is equal to
\begin{equation}
    \label{eq:decays-invisible}
    \Gamma(Z'\rightarrow\overline{\chi}{\chi})=\frac{\alpha_{D}m_{Z'}}{3}\bigg(1+\frac{2m_\chi^2}{m_{Z'}^2}\bigg)\sqrt{1-\frac{4m_\chi^2}{m_{Z'}^2}},
\end{equation}
with $\alpha_D=g_\chi^2/(4\pi)$. In the mass hierarchy below the resonance, $m_{Z'}<2m_\chi$, the $t-$channel annihilation is $\bar{\chi}\chi\rightarrow Z^\prime Z^\prime$, with $\langle\sigma v\rangle\propto g_{\chi}^4/m_\chi^2$.\\ \indent
Besides probing the particle nature of DM, this model could be relevant to the muon anomalous magnetic moment $(g-2)_\mu$. Within this framework, the discrepancy between the experimental \cite{Muong-2:2023cdq} and SM predicted \cite{Aoyama:2020ynm} $(g-2)_\mu$ values is explained through the loop correction \cite{Gninenko:2001hx,Pospelov:2007mp} $\Delta a_\mu^{Z'}=a_\mu(\text{Exp})-a_\mu(\text{SM})$ given by \cite{Gninenko:2014pea,Gninenko:2001hx,Chen:2017awl,Gninenko:2018tlp,Kirpichnikov:2020tcf,Amaral:2021rzw} 
\begin{equation}
	\label{eq:alpha-light-vector}
	\Delta a_\mu^{Z'}=\frac{g_{Z^{\prime}}^2}{4\pi^2}\int_{0}^{1}dx\ \frac{x^2(1-x)}{x^2+(1-x)m_{Z'}^2/m_\mu^2}.
\end{equation}
The current observables constraining the corresponding viable $m_{Z^\prime}$ upper mass bound arise from direct searches, sensitive to the kinematically allowed visible decay channel $Z^\prime\rightarrow\mu^{+}\mu^{-}$ \cite{BaBar:2016sci,CMS:2018yxg,ATLAS:2023vxg,Kamada:2015era,CCFR:1991lpl,CHARM-II:1990dvf}, restricting $m_{Z^\prime}<2m_\mu$. Constraints from invisible search, $Z'\rightarrow\bar{\chi}\chi$ can be found in, e.g., \cite{Belle-II:2022yaw}. The lower bound is set through the $Z'$ contribution to the radiation density of the Universe through $\Delta N_\text{eff}$, with its value being defined from both the CMB spectrum \cite{Planck:2018vyg} and the Big Bang nucleosynthesis (BBN) \cite{Ahlgren:2013wba,Kamada:2015era,Escudero:2019gzq} to $m_{Z'}>3-10$ MeV depending on the combination of the data \cite{Sabti:2019mhn}. Within this range, it is found that $g_{Z^\prime}\sim10^{-4}-10^{-3}$. \\ \indent
In this work, an approach to search for such a vector boson is presented, relying on the missing energy/momentum technique with the NA64 \cite{NA64:2016oww} beam  dump experiment at the CERN Super Proton Synchrotron (SPS). In particular, this paper relates in detail the methodology and findings of Ref. \cite{Andreev:2024sgn}. Besides, new results from both the search for a light spin$-0$ mediator and $A'\rightarrow\text{invisible}$ are discussed, together with expected projections for the forthcoming years. The paper is organized as follows. In Sec. \ref{sec:search-phenomenology}, the method of search is presented as well as its experimental signature. In addition, as a benchmark scenario, an overview of the underlying $Z'$ phenomenology is given, with particular emphasis on the differential and total production cross-sections, following the works of \cite{Kirpichnikov:2021jev,Sieber:2023nkq}. In this framework, the Weisz\"{a}cker-Williams phase-space approximation is used to simplify the computations. A detailed overview of the M2 beam line and the experimental set-up detectors is given in Sec. \ref{sec:na64mu-detectors}. Sec. \ref{sec:mc-approach} outlines the Monte Carlo (MC) framework for both signal event and beam optics simulations. The event reconstruction and detailed data analysis are presented in Sec. \ref{sec:data-analysis}, together with the validation of the aforementioned MC framework. Those are followed by a detailed discussion of the background sources in Sec. \ref{sec:background-sources}. The estimate of the signal yield, as well as its corresponding systematics, are discussed in respectively Secs. \ref{sec:signal-yield} and \ref{sec:systematics}. The final results of the search for invisibly decaying $Z'$ are presented in Sec. \ref{sec:results}, and new limits on both a muon-philic scalar particle and the invisibly decaying dark photon are shown. For all scenarios, the projected sensitivity to the expected statistics accumulated before and after the CERN LS3 is given. Finally, in Sec. \ref{sec:conclusion}, the main outcomes of this work are summarised.
\section{Method of search and model phenomenology\label{sec:search-phenomenology}}
\begin{figure}[H]
    \centering
    \includegraphics[width=0.4\textwidth]{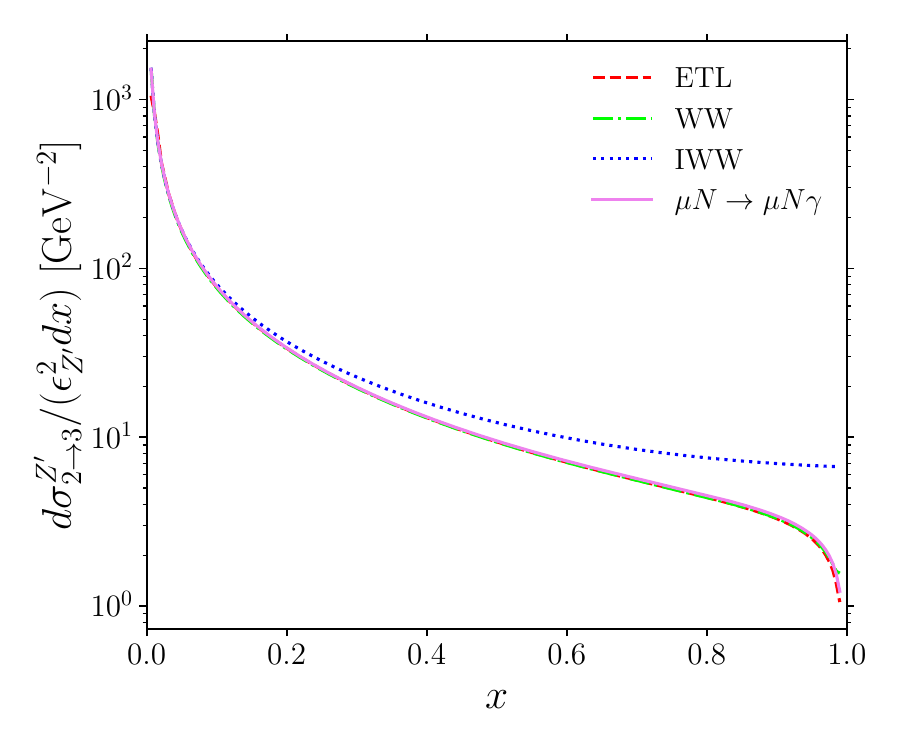}
    \caption{\label{fig:SM-DS-brem}Comparison of the single-differential cross-section as a function of the fractional energy of the emitted $\gamma$ and $Z'$ through respectively SM muon-bremsstrahlung and dark bremsstrahlung, $\mu N\rightarrow\mu NZ'$, in the mass limit $m_{Z'}\rightarrow0$. These results are obtained both at ETL and in the WW and IWW approaches, with mixing strength $\epsilon_{Z'}=g_{Z'}/e=1$. More details can be found in Ref. \cite{Kirpichnikov:2021jev}.}
\end{figure}

The $Z'$ vector boson could be produced in the bremsstrahlung-like reaction of a muon interacting with a target's nucleus, $N$. In particular, as discussed in the previous section, for sufficiently small coupling values of $g_{Z'}$, or within the kinematic limit set by $m_{Z'}\leq2m_\mu$, the light mediator predominantly decays invisibly, $Z'\rightarrow\text{invisible}$.\\ \indent
The method of search is described in details in \cite{Gninenko:2014pea,NA64:2018iqr,Sieber:2021fue}. Should a light $Z'$ vector boson exists, it could be produced in the bremsstrahlung process of highly energetic muons, with incoming energy $E_{0}$, impinging on an active target
\begin{equation}
    \label{eq:production}
    \mu N\rightarrow\mu NZ';\ Z'\rightarrow\text{invisible}.
\end{equation}
The $Z'$ then carries away a fraction $x$ of the muon beam energy, $xE_{0}$, and either promptly decays invisibly or propagates downstream of the target without interaction in the sub-detectors, where the outgoing scattered muon is detected with remaining energy $(1-x)E_{0}=yE_0$. \\ \indent
The occurence of a $Z'$ signal event produced through the reaction of Eq. \eqref{eq:production} would appear as a single scattered muon with momentum about less than half that of its initial one, no activity in the VETO detectors downstream of the target, an energy deposition in the calorimeters compatible with that of a minimum ionising particle (MIP) and a single reconstructed particle track past the interaction point.\\ \indent
The production cross-section associated with the process of Eq. \eqref{eq:production} is computed in detail in previous works \cite{Kirpichnikov:2021jev,Sieber:2023nkq}, following the kinematics $\mu(p)+N(P_i)\rightarrow\mu(p')+N(P_f)+Z'(k)$. Both the exact tree-level (ETL) and the Weiszäcker-Williams (WW) phase space approximation are covered. The computations are carried out under the assumptions that the nucleus has zero spin and that the contribution to the nucleus-photon vertex factor is purely elastic. This implies neglecting the inelastic effects due to the high atomic number $Z$ of the lead (Pb) target material, with $Z=82$ and atomic weight $A=207$. As such, it takes the form \cite{Kirpichnikov:2021jev}
\begin{equation}
    ieF(t)\mathcal{P}_\mu,\ F^2(t)\equiv G_\text{el}(t)=Z^2\bigg(\frac{a^2t}{1+a^2t}\bigg)^2\bigg(\frac{1}{1+t/d}\bigg),
\end{equation}
with $\mathcal{P}_\mu=(P_i-P_f)_\mu$ the four-momentum transfer associated with the nucleus, $a=111Z^{-1/3}/m_e$ and $d=0.164A^{-2/3}\text{GeV}^2$ screening effect factors related to respectively the Coulomb field from atomic electrons and the nucleus size. In the WW approach, the $2\rightarrow3$ production process depicted in Eq. \eqref{eq:production} and given by the ETL is factorised through the equivalent photon flux (WW) approximation \cite{Liu:2016mqv,Liu:2017htz} into a $2\rightarrow2$ process given the virtual photon flux
\begin{equation}
    \label{eq:photon-flux}
    \chi^\text{WW}=\int_{t_\text{min}}^{t_\text{max}}dt\ \frac{t-t_\text{min}}{t^2}F^2(t),
\end{equation}
with $t_\text{min}$ and $t_\text{max}$ the minimum and maximum momentum transfer to the nucleus given in, e.g. \cite{Kirpichnikov:2021jev}. This leads to the following simplification of the double-differential cross-section
\begin{equation}
\label{eq:WWxtheta}
\frac{d^2 \sigma^{Z'}_{2\to 3}}{d x d \cos\theta_{Z'}} \Bigr|_\text{WW} \simeq \frac{\alpha \chi^\text{WW}}{ \pi (1-x)} E_0^2 x \beta_{Z'}  \frac{d \sigma^{Z'}_{2\to 2} }{d (pk)} \Bigr|_{t=t_\text{min}},
\end{equation}
with $\theta_{Z'}$ the emission angle of $Z'$, $\alpha$ the fine structure constant, $\beta_{Z'}=k/m_{Z'}$ the $\beta$-factor associated with the $Z'$. It is worth noting that in the case where $t_\text{max/min}$ do not depend on $(x,\ \theta_{Z'})$, Eq. \eqref{eq:photon-flux} further simplifies and Eq. \eqref{eq:WWxtheta} is computed through the so-called \emph{improved} WW (IWW) approximation at the cost of loosing accuracy. Because the method of search relies on the final-state muon kinematics, Eq. \eqref{eq:WWxtheta} is also computed for the $(y,\ \psi_\mu^\prime)$ tuple of variables, with $\psi_\mu^\prime$ being defined as the scattered muon emission angle.
\begin{figure}[H]
    \centering
    \includegraphics[width=0.4\textwidth]{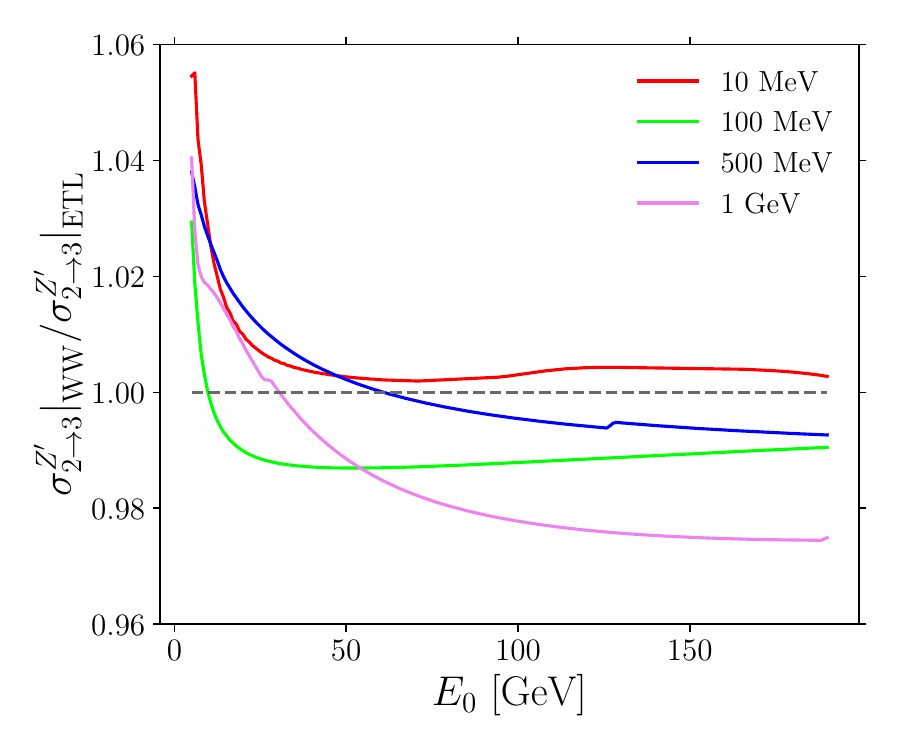}
    \caption{\label{fig:total-cs}Ratio of the cross-sections computed in the WW approximation and at ETL, $\sigma_{2\rightarrow3}^{Z'}\big|_\text{WW}/\sigma_{2\rightarrow3}^{Z'}\big|_\text{ETL}$, as a function of the muon beam energy, $E_0$, for different $Z'$ masses, $m_{Z'}$. More details can be found in \cite{Kirpichnikov:2021jev}.}
\end{figure}
Within the kinematic regime of interest for the NA64$\mu$ experiment, given $E_{0}\gg m_\mu,\ m_{Z'}$, the WW approximation reproduces accurately the ETL, with relative error defined as $(\mathcal{O}_\text{ETL}-\mathcal{O}_\text{WW})/\mathcal{O}_\text{ETL}$ of the order of $\leq5\%$ over the mass range $m_{Z'}\in[10\ \text{MeV},\ 1 \text{GeV}]$, for $\mathcal{O}$ being the observables associated with the single-differential and total cross-sections \cite{Kirpichnikov:2021jev,Sieber:2023nkq}. For illustrative purposes, the ratio of the total production cross-sections $\sigma_{2\rightarrow3}^{Z'}$ evaluated within the WW approximation and at ETL, as a function of the beam energy, $E_{0}$, is shown in Fig. \ref{fig:total-cs}. In the lower energy regime, $E_{0}<10$ GeV, the error is the largest, especially for $m_{Z'}=1$ GeV, due to the assumption $E_{0}\gg m_\mu,\ m_{Z'}$.\\ \indent
The computational framework is further validated in the zero mass limit, $m_{Z'}\rightarrow0$, against SM muon bremsstrahlung, $\mu N\rightarrow\mu N\gamma$, as expressed in \cite{Kelner:1995hu,Kelner:1997cy,Groom:2001kq}. The dependence of the single-differential cross-section on the fractional energy is shown in Fig. \ref{fig:SM-DS-brem}. The dark bremsstrahlung cross-section reproduces SM events with photon emission with a relative error $\leq2\%$ in both the ETL and WW frameworks, but exceed $50\%$ for $x\rightarrow1$ in the IWW approach, due to the simplification of the flux term's boundaries of Eq. \eqref{eq:photon-flux} \cite{Kirpichnikov:2021jev}.
These computations lay the basis for implementing the model within a computer-based program to perform a realistic MC simulation and signal propagation study as discussed in Sec. \ref{subsec:signal-production}. 
\section{The M2 beam line and NA64$\mu$ detectors\label{sec:na64mu-detectors}}
The experiment employs the unique M2 beam line \cite{Doble:1994np} at the CERN North Area (NA), delivering to the experimental hall north 2 (EHN2) high-intensity muons up to $2\times10^{8}$ $\mu/$spill, mostly in the energy range $100-225$ GeV. Those are produced by highly energetic $\sim450$ GeV/c protons extracted from the CERN SPS with a maximum flux of $1.5\times10^{13}$ protons per SPS cycle \cite{Banerjee:2774716}, and impinging on a thick Beryllium (Be) target, mostly producing pions and kaons, $p+\text{Be}\rightarrow\pi, K+X$. These secondary hadrons predominantly decay to muons along a 600-meter-long decay section, while those surviving are suppressed by a series of 2.7 interaction length ($\lambda_\text{int}$) Be absorbers yielding a level of hadron admixture $\pi/\mu$ of $10^{-6}-10^{-5}$ with $K/\pi\sim0.3$ \cite{Doble:1994np}. The final muon beam optics for NA64$\mu$ are defined through a series of quadrupoles in a focussing-defocussing (FODO) scheme, as well as magnetic collimators, defining the \emph{muon section}, and resulting in muons mostly within the momentum band $\sim160\pm3$ GeV/c, with beam spot divergency $\sigma_x\sim0.9$ cm and $\sigma_y\sim1.9$ cm \, \cite{Sieber:2021fue} at the entrance of the experimental set-up shown in Fig. \ref{fig:experimental-set-up-upstream}. This choice of energy range results from detailed MC simulation to optimize the hermeticity of the set-up given an optimal beam intensity.\\ \indent
As highlighted in Sec. \ref{sec:search-phenomenology}, the method of search for a signal event at NA64$\mu$ must rely on a precise knowledge of both the incoming, $p_\text{in}$, and outgoing, $p_\text{out}$, muon momenta given the signature $Z'\rightarrow\text{invisible}$. The different parameters of the detectors used for the 2022 muon run are summarized in Table \ref{tab:detectors_parameters}.\\ \indent
 The initial momentum is measured along part of the muon section utilizing the existing MBN-type dipole vertical bending magnets (BEND6), each with length $\simeq5$ meters and $B\cdot L=5$ T$\cdot$m, and a set of six scintillator hodoscopes, the beam momentum stations (BMS$_{1-6}$), with a spatial and time resolution of respectively $\sigma_s\simeq0.4-2.5$ mm and $\sigma_t\simeq0.3-0.5$ ns \cite{EuropeanMuon:1980nje,COMPASS:2007rjf}. This first magnet spectrometer, MS1, is completed by a series of four Micro-Mesh Gaseous (Micromegas) detectors (MM$_{1-4}$) and two straw-tube chambers (ST$_{5,4}$), which form the \emph{upstream} part of the experimental set-up (see Fig. \ref{fig:experimental-set-up-upstream}). Those tracking detectors have respectively a spatial and time resolution of $\sigma_s\simeq100\ \mu$m, $\sigma_t\simeq15$ ns \cite{Banerjee:2017mdu} and $\sigma_s\simeq1$ mm, $\sigma_t\simeq5$ ns \cite{Volkov:2019qhb}. The resulting achieved resolution on the momentum is $\sigma_{p_\text{in}}/p_\text{in}\simeq3.8\%$.\\ \indent Given a well-defined incoming muon, the signal event is expected to be produced within an active Shashlik-like electromagnetic calorimeter (ECAL). The detector has a 150-layer longitudinal segmentation, with respectively 1.5 mm thick lead (Pb) and 1.5 mm plastic scintillator (Sc) plates for a total radiation length of $40X_0$ ($X_0(\text{Pb})=0.56$ cm), among which $4X_0$ are used as pre-shower detector. The lateral granularity of the ECAL module is given by a $5\times6$ matrix of cells, each with a width and height of $\simeq3.83$ cm. The light from the scintillator plates is collected through 1 mm wavelength shifter (WLS) fibers, achieving an energy resolution of $\sigma_{E}/E=8\%/\sqrt{E}\bigoplus1\%$. Following the ECAL, a large $55\times55$ cm$^2$ veto counter (VETO) made of three 5 cm thick scintillators stacked together is used to reject upstream interactions from the target, with a minimum ionizing particle (MIP) inefficiency estimated from muon data to be $\leq10^{-3}$. To enforce the rejection of non MIP-compatible events, the veto system is completed by a 100-cm-long prototype sampling hadronic calorimeter (VHCAL), with a 30-layer segmentation, each layer being made of 25 mm thick copper (Cu) and 2 mm thick plastic scintillator plates. The VHCAL cross-section is defined by a $4\times4$ matrix of cells, each with a $12\times12$ cm$^2$ area, with the peculiarity of having a $12\times6$ cm$^2$ hole in its middle to reject events with large angle charged secondaries and identify single final-state muon events passing through the detector without activity. The total stopping power of the module is $5\lambda_\text{int}$. The optimization of the VHCAL is an ongoing activity to provide a full-scale detector aiming at improving the hermeticity of the set-up required after the CERN Long Shutdown3 (LS3) (see Sec. \ref{sec:results}).\\ \indent
The outgoing muon momentum is reconstructed through a second magnet spectrometer, MS2, utilizing an MBP-type horizontal dipole magnet, with maximum bending power $B\cdot L=3.8$ T$\cdot$m and with a length of $\simeq2$ meters. Three Micromegas (MM$_{5-7}$), two straw chambers (ST$_{2-1}$) and four Gaseous Electron Multiplier (GEM$_{1-4}$) trackers \cite{Ketzer:2004jk} sandwich the bending magnet to achieve a momentum resolution of $\sigma_{p_\text{out}}/p_\text{out}\simeq4.4\%$. Similarly to the MM detectors, the GEMs achieve a spatial and time resolution of $\sigma_s\simeq115\ \mu$m and $\sigma_t\simeq\mathcal{O}(10\ \text{ns})$.\\ \indent
The final-state muons are identified by two large hadronic calorimeter (HCAL$_{1,2}$) modules. Each of them is a $\sim154$-cm-long sampling module, longitudinally segmented with 48 layers of 25 mm thick steel (Fe) and 4 mm plastic scintillator plates, separated by a 9-mm-air gap for a total interaction length of $7.5\lambda_\text{int}$. To ensure maximal hermeticity along the deflection direction, a single module has a lateral segmentation given by $6\times3$ matrix with cell dimensions $194\times192$ cm$^2$. Similarly, as for the ECAL module, the readout is performed through WLS fibers connected to the Sc plates toward the photomultipliers (PMTs), thus achieving an energy resolution of $\sigma_E/E=65\%/\sqrt{E}+6\%$. In addition to particle identification (PID) through energy deposit in the HCAL, a large $120\times60$ cm$^2$ UV straw chamber (ST$_{11}$) is located at the end of the second HCAL module, HCAL$_2$. The aforementioned spectrometer and detectors build the \emph{downstream} part of the experiment (see Fig. \ref{fig:experimental-set-up-downstream}).\\ \indent
The trigger system is defined by a set of two 42-mm-diameter, $\sim8\times10^{-3}X_0$ thick, scintillator counters (S$_0$ and S$_1$) together with a $10\times10$ cm$^{2}$, $\sim5\times10^{-4}X_0$ thick, veto counter (V$_0$) with a 45-mm-diameter hole in its middle. This allows for a clean selection of muons from the core of the beam \cite{Gninenko:2014pea,Sieber:2021fue} and defines the \emph{calibration} trigger configuration ($\text{S}_0\times\text{S}_1\times\overline{\text{V}_0}$) coping with an intensity as high as $2.8\times10^{6}$ $\mu/$spill. The system is completed by two large rectangular $20\times20$ cm$^2$ scintillator counters (S$_4$ and S$_\mu$) to trigger on the deflected muons after MS2, with a relative measured trigger rate of $\mathcal{O}(10^{-4})$ that of the calibration trigger coincidence. As such, this defines the \emph{physical} trigger configuration with\\ \indent
\begin{equation}
    \label{eq:trigger-physical}
    \text{Trig.}=\text{S}_0\times\text{S}_1\times\overline{\text{V}_0}\times\text{S}_4\times\text{S}_\mu,
\end{equation}
designed to accept events that hit in-time the scintillator counters, such that $\delta_t\leq 3$ ns for the coincidence gate $\text{S}_0\times\text{S}_1\times\overline{\text{V}_0}$ and $\delta_t\leq20$ ns for the full configuration of Eq. \eqref{eq:trigger-physical}.
\begin{table}[H]
    \centering
    \resizebox{0.48\textwidth}{!}{
    \begin{tabular}{llll}
        \hline
        \hline
        Element & $N$ & Active area (cm$^2$)  & Resolution \\
        \hline
        BMS1-4 & 4 & $(6-12)\times(9-23)$ & $\sigma_s=1.3-2.5$ mm\\ 
        & & & $\sigma_t=0.3$ ns \\
        BMS5 & 1 & $12\times16$ & $\sigma_s=0.7$ mm \\
        & & & $\sigma_t=0.5$ ns \\
        BMS6 & 1 & $12\times16$ & $\sigma_s=0.4$ mm \\
        & & & $\sigma_t=0.5$ ns \\
        MM1-4 & 4 & $8\times8$ & $\sigma_s\simeq100\ \mu$m\\
        & & & $\sigma_t\simeq15$ ns \\
        ST4-5 & 2 & $20\times20$ & $\sigma_s\simeq1$ mm\\
        & & & $\sigma_t\simeq5$ ns \\
        ECAL & 1 & $19\times23$ & $\sigma_E/E=8\%/\sqrt{E}\bigoplus1\%$ \\
        VETO & 1 & $55\times55$ & $\sigma_E/E=3\%$ at 1 MeV \\
        & & & $\sigma_t\simeq2$ ns \\
        VHCAL & 1 & $50\times50$ & $\sigma_E/E=45\%/\sqrt{E}+5\%$ \\
        GEM1-4 & 4 & $10\times10$ & $\sigma_s\simeq115\ \mu$m\\
        & & & $\sigma_t\simeq10$ ns \\
        ST1-2 & 2 & $20\times20$ & $\sigma_s\simeq1$ mm \\
        & & & $\sigma_t\simeq5$ ns \\
        MM5-7 & 3 & $25\times80$ & $\sigma_s\simeq100\ \mu$m \\
        & & & $\sigma_t\simeq15$ ns \\
        HCAL & 2 & $120\times60$ & $\sigma_E/E=65\%/\sqrt{E}+6\%$ \\
        S$_{0}$, S$_{1}$ & 2 & 42 mm & $\sigma_t\simeq1$ ns \\
        S$_{4}$, S$_{\mu}$ & 2 & $20\times20$ & $\sigma_t\simeq1$ ns \\
        V$_{0}$ & 1 & $10\times10$ & $\sigma_t\simeq1$ ns \\
        \hline
        \hline
    \end{tabular}
    }
    \caption{Summary of the detector elements for the 2022 muon run NA64 experimental set-up. The first column refers to the name of the detector, the second to the number of elements within the set-up. The third columns gives the active area of such a detector in cm$^2$. In the fourth column, $\sigma_s$, $\sigma_t$ and $\sigma_E/E$ denote respectively the spatial, time and energy resolution of the element.}
    \label{tab:detectors_parameters}
\end{table}
\begin{widetext}
\par\smallskip\noindent
\centerline{\begin{minipage}{\linewidth} 
\begin{figure}[H]
    \centering
    \includegraphics[width=0.99\linewidth]{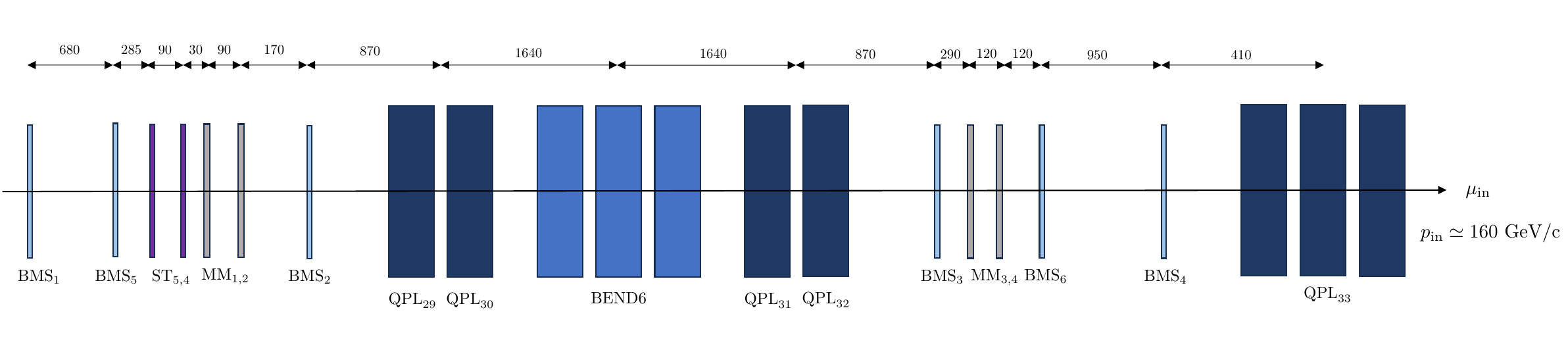}
    \caption{\label{fig:experimental-set-up-upstream}Experimental set-up schematic overview of the search for $Z^\prime\rightarrow\text{invisible}$ \cite{Andreev:2024sgn}. Top: The \emph{upstream} experimental region for the reconstruction of the incoming muon momentum through the MS1 (BEND6) magnet spectrometer using MM$_{1-4}$, ST$_{5,4}$ and BMS$_{1-6}$. The beam-defining optics quadrupoles, QPL$_{29-32}$ and QPL$_{33}$, part of the FODO scheme, are shown. For completeness, the distances between the different detector elements are given in cm. See text for more details.}
\end{figure}
\end{minipage}}
\end{widetext}
\begin{widetext}
\par\smallskip\noindent
\centerline{\begin{minipage}{\linewidth} 
\begin{figure}[H]
    \centering
    \includegraphics[width=0.99\linewidth]{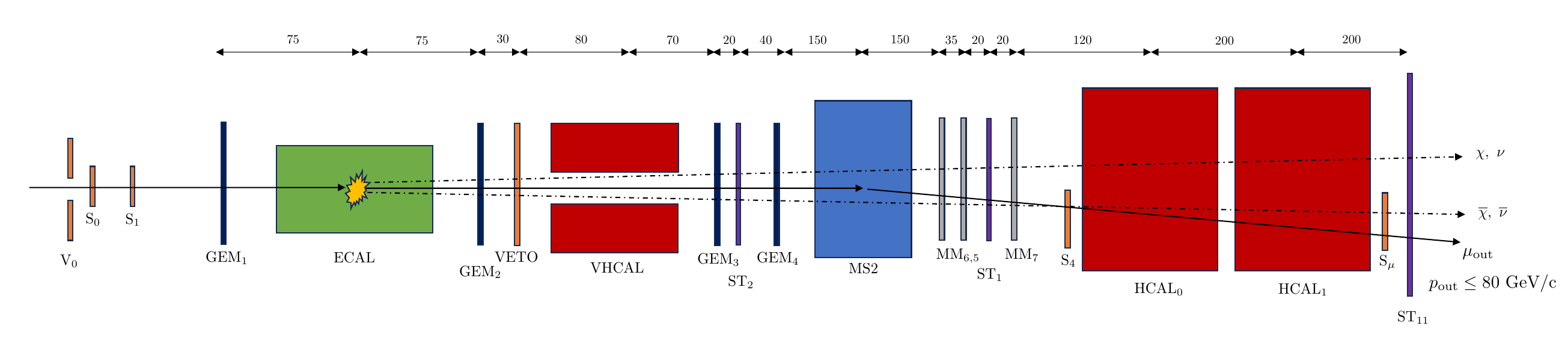}
    \caption{\label{fig:experimental-set-up-downstream}The \emph{downstream} region of the experiment for final-state muon identification through detector response and momentum reconstruction in MS2. The distances between the detector elements are given in cm. See text for more details.}
\end{figure}
\end{minipage}}
\end{widetext}
\section{Monte Carlo approach\label{sec:mc-approach}}
\subsection{Signal event production\label{subsec:signal-production}}
\begin{figure}[H]
    \centering
    \includegraphics[width=0.4\textwidth]{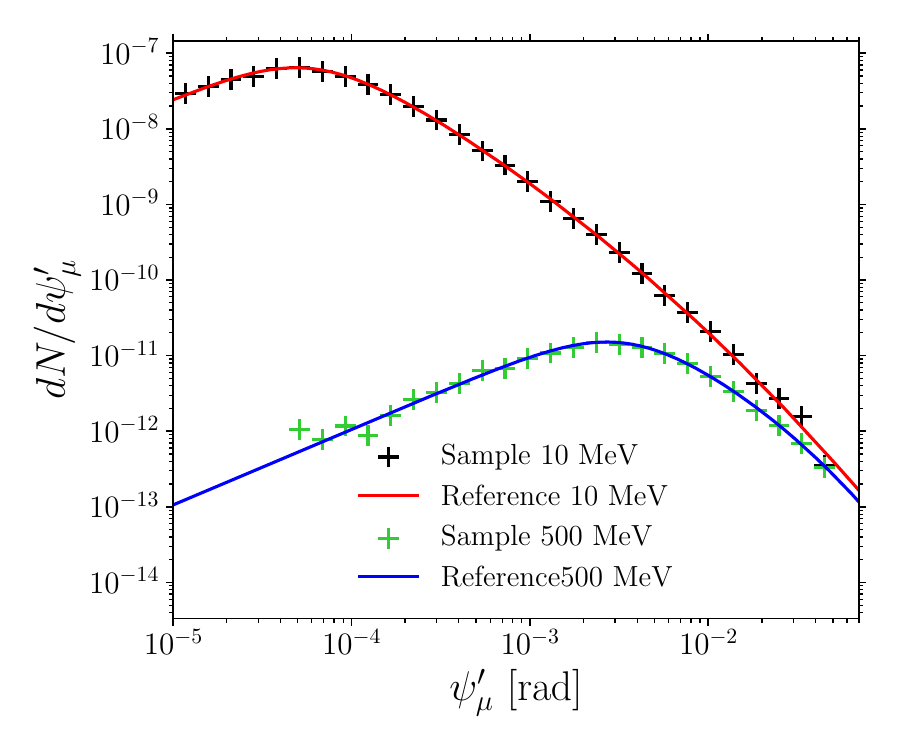}
    \caption{\label{fig:dmg4-sample-10MeV}The muon emission angles $\psi_{\mu}^\prime$ for kinematical regimes with $m_{Z'}=10,500$ MeV. The normalized sampled angle (crosses) and the normalized target partial distribution function (PDF, lines) are shown for comparison. Small deviations from the target PDF at larger $\psi_\mu^\prime$ are due to fewer statistics in the binned sample distribution.}
\end{figure}

The signal is simulated within the \texttt{GEANT4}-based \cite{GEANT4:2002zbu,Allison:2016lfl} application programming interface (API) \texttt{DMG4} toolkit \cite{Bondi:2021nfp,Kirsanov:2023evm,Oberhauser:2024ozf}. The number of $Z'$ produced at each step $k$ within the target is given through
\begin{equation}
    \mathcal{N}_{Z'}^{(k)}=\frac{\rho\mathcal{N}_{A}}{A}\sigma_{2\rightarrow3}^{Z'}\Bigr|_\text{WW}(E_k)\Delta l_k,
\end{equation}
with $\rho$ and $A$ respectively the density and atomic weight of the lead ECAL target, $\mathcal{N}_{A}\simeq6.022\times10^{-23}$ mol$^{-1}$ the Avogadro's number, $\sigma_{2\rightarrow3}^{Z'}\Bigr|_\text{WW}(E_k)$ the total production cross-section for the $Z'$ boson discussed in Sec. \ref{sec:search-phenomenology} for a muon with energy $E_k$ and $\Delta l_k$ the corresponding step length.\\ \indent
During this phase, a signal event is simulated as follows: (i) The interaction probability is inferred from run-time tabulated cross-sections values through the mean free path associated with the production process. (ii) Shall a $Z'$ vector boson be produced, its phase-space is sampled from the single- and double-differential cross-sections $d^2\sigma_{2\rightarrow3}^{Z'}/dxd\theta_{Z'}$ and $d\sigma_{2\rightarrow3}^{Z'}/dx$ through a direct Von Neumann accept/reject scheme \cite{UBHD66609337}. The final state's muon kinematics are similarly treated for the variables $y$ and $\psi_\mu^\prime$. (iii) Finally, depending on the selected decay channels and values of $(m_{Z'},\ g_{Z'})$, the newly produced particle can decay within the set-up.\\ \indent
The computer-based program implementation of the production cross-sections discussed in Sec. \ref{sec:search-phenomenology} are done within the WW approach. For a decrease of the computing time, and thus an improved run-time performance, the analytical expressions of Ref. \cite{Sieber:2023nkq} are adopted by integrating out the emission angle $\theta_{Z'}$ of Eq. \eqref{eq:WWxtheta} to obtain an exact form for $d\sigma_{2\rightarrow3}^{Z'}/dx$, such that the total production cross-section reads
\begin{equation}
\label{eq:analytical-muon}
\begin{split}
&\sigma_{2\rightarrow 3}^{Z'}\Bigr|_\text{WW}=\\
&\int_{x_\text{min}}^{x_\text{max}}dx\ \bigg(\epsilon_{Z'}^2\alpha^3\sqrt{1-\frac{m_{Z'}^{2}}{E_0^2}}\frac{1-x}{x}\sum_{i=1}^{6}I_i^{Z'}(x,\tilde{u})\bigg\rvert_{\tilde{u}=\tilde{u}_\text{min}}^{\tilde{u}=\tilde{u}_\text{max}}\bigg),
\end{split}
\end{equation}
where the six special functions $I_i^{Z^\prime}$, $i=1,2,...6$ and the definition of the variable $\tilde{u}_\text{min}$ and $\tilde{u}_\text{max}$ can be found in \cite{Sieber:2023nkq}. As discussed in Sec. \ref{sec:search-phenomenology}, the accuracy of the method is inferred by comparing the absolute yield of the photon and massless $Z'$ bremsstrahlung productions. In this optic, events are extracted from a \texttt{GEANT4} simulation of the NA64$\mu$ ECAL target. For an appreciable comparison, an adequate choice of production cut (threshold), set to 1 GeV, similar values of $t_\text{min}$ and $t_\text{max}$ within the two processes, and equal parameters $x_\text{min}$ and $x_\text{max}$ (respectively $v_\text{cut}$ and $v_\text{max}$ within \texttt{GEANT4}, see \cite{Bogdanov:2006kr}) are chosen. The resulting single-differential distributions are shown in Fig. \ref{fig:dmg4-g4-bremsstrahlung}, left, for the events' sampled fractional energies, $x$, assuming a muon beam with mono-energetic energy $E_{0}=160$ GeV. \\ \indent
This procedure is repeated for several values of $E_{0}$, and the resulting distributions $dN_{2\rightarrow3}/dx$ integrated to compute the absolute $\gamma$ and $Z'$ yields. Over the full beam energy range, it is found that the ratio of the yields, $N_{Z'}/N_{\gamma}$, is on average $\sim0.98\pm0.01$. The corresponding extracted values of the production cross-sections are shown in Fig. \ref{fig:dmg4-g4-bremsstrahlung}, right.
\begin{widetext}
\par\smallskip\noindent
\centerline{\begin{minipage}{\linewidth} 
\begin{figure}[H]
    \centering
    \includegraphics[width=0.37\textwidth]{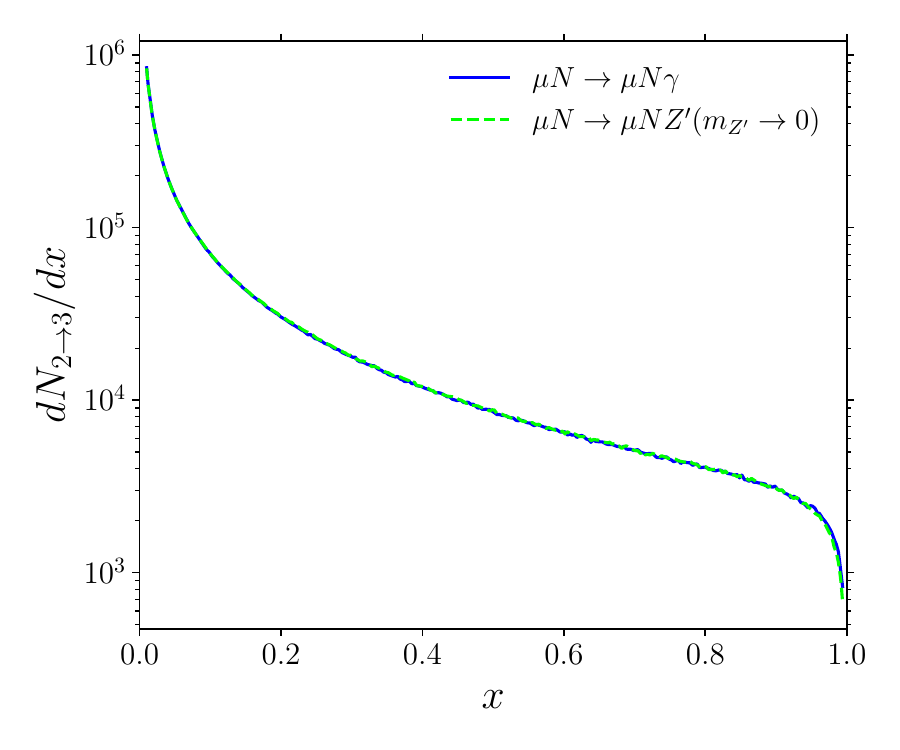}
    \hspace{1cm}
    \includegraphics[width=0.37\textwidth]{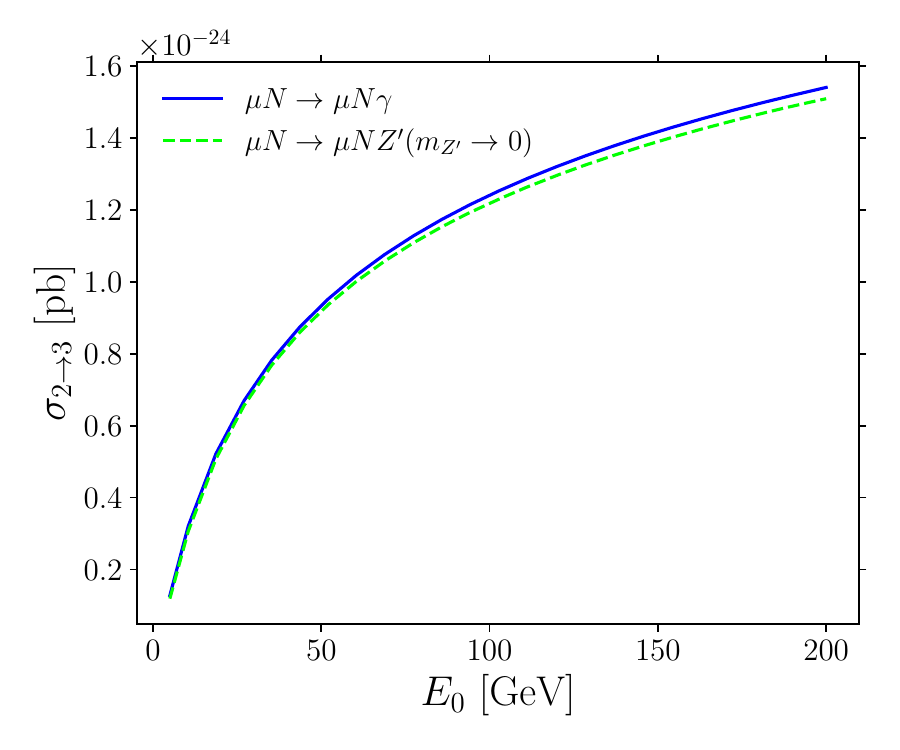}
    \caption{\label{fig:dmg4-g4-bremsstrahlung}Left: Distributions of the fractional energy, $x$, for repsectively SM muon bremsstrahlung, $\mu N\rightarrow\mu N\gamma$ and dark bremsstrahlung in the limit $m_{Z'}\rightarrow0$. The events are obtained from a minimal \texttt{GEANT4} simulation of the NA64$\mu$ target, assuming a fixed muon beam energy $E_0=160$ GeV. Right: Production cross-sections $\sigma_{2\rightarrow3}$ as a function of the muon beam energy, $E_0$, extracted from a realistic \texttt{GEANT4} simulation of both SM muon bremsstrahlung, $\mu N\rightarrow\mu N\gamma$, and dark bremsstrahlung in the limit $m_{Z'}\rightarrow0$, within the NA64$\mu$ ECAL.}
\end{figure}
\end{minipage}}
\end{widetext}
Because the experimental signature strongly relies on the scattered muon kinematics, the sampling of the final state emission angle is implemented based on the WW approximation of $d^2\sigma_{2\rightarrow3}^{Z'}/dyd\psi_{\mu}^\prime$, which reads \cite{Kirpichnikov:2021jev}
\begin{equation}
\label{eq:WWypsi}
\frac{d^2 \sigma^{Z'}_{2\to 3}}{d y d \cos\psi_{\mu}^\prime} \Bigr|_\text{WW} \simeq \frac{\alpha \chi^\text{WW}}{ \pi (1-y)} E_0^2 y \beta_{\mu}^\prime  \frac{d \sigma^{X}_{2\to 2} }{d (pp')} \Bigr|_{t=t_\text{min}},
\end{equation}
with $y$ the final-state muon fractional energy, $\psi_\mu^\prime$ its emission angle, $\beta_\mu^\prime$ its Lorentz $\beta-$factor and $p^\prime$ its four-momenta. The accuracy in reproducing the underlying probability distribution function (PDF) is inferred through a Kolmogorov-Smirnov (KS) statistical test for goodness of the fit \cite{Kolmogorov:1933sde,Smirnov:1948tfe}. In Fig. \ref{fig:dmg4-sample-10MeV} is shown the sampled values of the muon scattered angles $\psi_\mu^\prime$ for a mass $m_{Z'}=10$ MeV.
\subsection{Beam optics simulations and trigger configuration optimisation\label{subsec:beam-optics}}
\begin{figure}[H]
    \centering
    \includegraphics[width=0.4\textwidth]{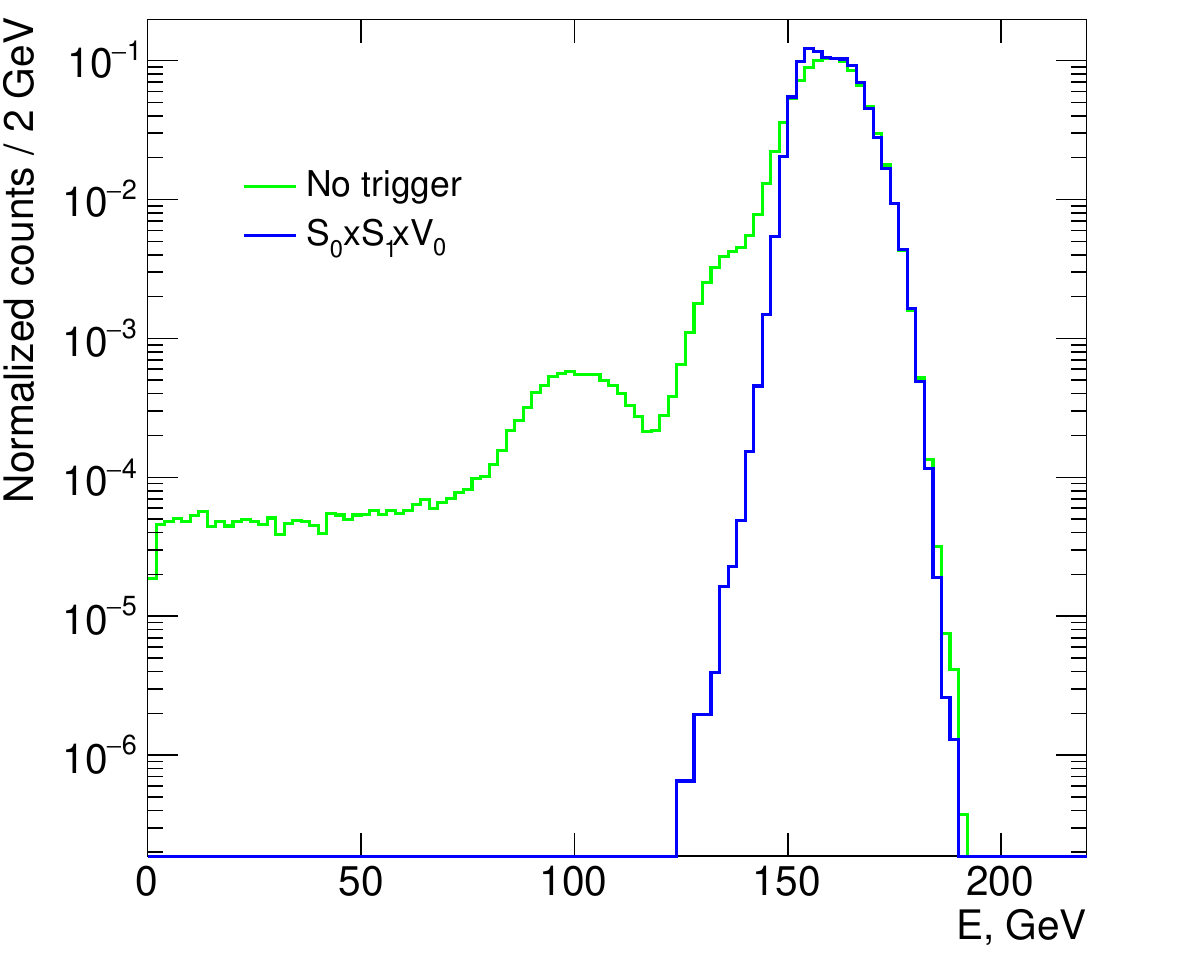}
    \caption{Calibration trigger acceptance effects (blue solid line) on the initial-state beam muons before the ECAL target.}
    \label{fig:trigger-calibration}
\end{figure}

\begin{figure}[H]
    \centering
    \includegraphics[width=0.4\textwidth]{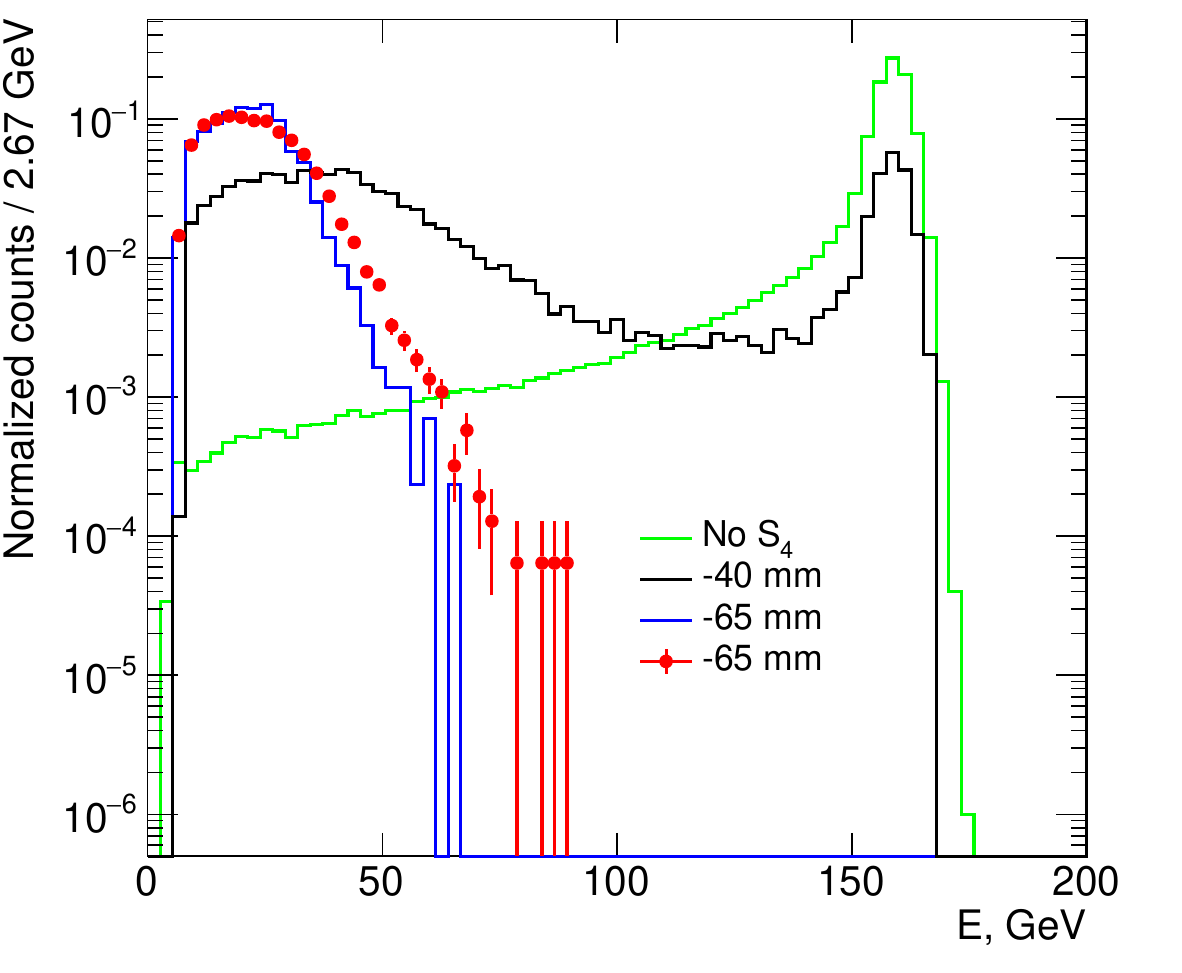}
    \caption{Final-state muons energy distribution from the SM process $\mu N\rightarrow\mu N\gamma$ using the calibration trigger configuration (green line) and different physical trigger configurations with the S$_4$ counter shifted respectively -40 mm (black line) and -65 mm (blue line) along the deflection axis past MS2. For completeness, bremsstrahlung-like events $\mu N\rightarrow\mu NZ'$ in the $m_{Z'}=1$ MeV scenario (solid red dots) are shown for S$_4$ at -65 mm.}
    \label{fig:trigger-physical}
\end{figure}

The full NA64$\mu$ beam optics are simulated through the \texttt{TURTLE} \cite{Brown:1974ns}, \texttt{TRANSPORT} \cite{Brown:1983jnh} and \texttt{HALO} \cite{Iselin:1974fu} programs as to transport the muon particles up to the first BMS detector (BMS$_{1}$), reproducing accurately the momenta and spatial distributions of both the core of the beam and its halo that constitutes about $20\%$ of the full beam intensity \cite{Sieber:2021fue}. Additionally, the beam particle composition is included based on the beam delivery simulation (\texttt{BDSIM}) \texttt{GEANT4} API \cite{Nevay:2018xwd,Nevay:2018zhp,Nevay:2019kmu,Persson:2021fas}, allowing secondaries produced in interactions with the upstream beam line material to be propagated towards the set-up. The above lays the basis for precisely propagating muons within the set-up and building the trigger system.\\ \indent
To efficiently remove the halo component of the beam being associated with low-momenta scattered muons being transported in the return fields of the upstream magnet yokes, the geometry of S$_0$, S$_1$ and V$_0$ is optimized to maximize the number of triggers on the core of the beam \cite{Sieber:2021fue} using the calibration trigger configuration (beam scintilaltor counters coincidences)). For 42-mm-diameter counters, $(57\pm3)\%$ of the MOT pass the trigger condition, while most of the lower-energetic component of the beam is rejected (see Fig. \ref{fig:trigger-calibration}).
The second part of the trigger system introduced in Sec. \ref{sec:na64mu-detectors} is optimized under several constraints, namely (i) an appreciable DAQ performance (limited to 10 kHz), maximizing (ii) the beam intensity ($\mu$/spill) and (iii) the number of triggers on produced signal events, scaling as $\sigma_{2\rightarrow3}^{Z'}\sim g_{Z'}^2\alpha^2Z^2/m_{Z'}^2$. To maximize the rejection of non-signal events, referred to as \emph{beam muons}, the typical trajectory of 160 GeV/c muons compatible with a MIP in the ECAL is studied within the set-up using a \texttt{GenFit}-based \cite{Rauch:2014wta} Runge-Kutta track extrapolator, accounting for the proper initial muon momentum and angular distributions. At the position of S$_4$ along the beam line, it is found that the average deflection past MS2 is $\langle\delta x\rangle\simeq-12$ mm. Based on this result, the trigger counters' positions and dimensions are optimized considering also the typical final-state signal muon emission angle scaling as $\psi_\mu^\prime\sim m_{Z'}/E_0$ \cite{Kirpichnikov:2021jev}, thus compensating the low production yield at high masses. In particular, for an appreciable comparison with the available data (see Sec. \ref{sec:data-analysis}), the MIP-compatible muons in the ECAL are compared against single hard-bremsstrahlung muon events in this detector, given by the reaction $\mu N\rightarrow\mu N\gamma$, as this mimics the final-state kinematics of signal muons (see Secs. \ref{sec:search-phenomenology} and \ref{subsec:signal-production}). A sample of muons from this reaction is simulated with \texttt{GEANT4} in the downstream part of the experimental set-up described in Sec. \ref{sec:na64mu-detectors}, requiring a single-photon emission in the ECAL from impinging muons. Fig. \ref{fig:trigger-physical} shows the energy distributions of final-state muons from both reactions \eqref{eq:production} and $\mu N\rightarrow\mu N\gamma$ under different physical trigger configurations, with for the later case also the distribution under the trigger condition $\text{S}_0\times\text{S}_1\times\overline{\text{V}_0}$. For a value of S$_4$ shifted 65 mm and S$_\mu$ shifted 152 mm along the deflection axis past MS2, it is found that the acceptance of signal candidates is optimized for the mass range 10 MeV to 1 GeV, with a scattered muon angle peaking around $\psi_\mu^\prime\sim10^{-2}$ rad. The effect on beam muon rejection is inferred by simulating, in addition to muon bremsstrahlung, the whole class of muon-induced electromagnetic and hadronic processes within the target with a single muon among the final states. From simulation, the corresponding physical trigger rate is found to be $(0.026\pm0.004)\%$ that of the calibration trigger, in good agreement with the results of Sec. \ref{sec:data-analysis}, and satisfying the constraints (i-ii).
\subsection{Detector hermeticity and optimization}
\begin{figure}[H]
    \centering
    \includegraphics[width=0.4\textwidth]{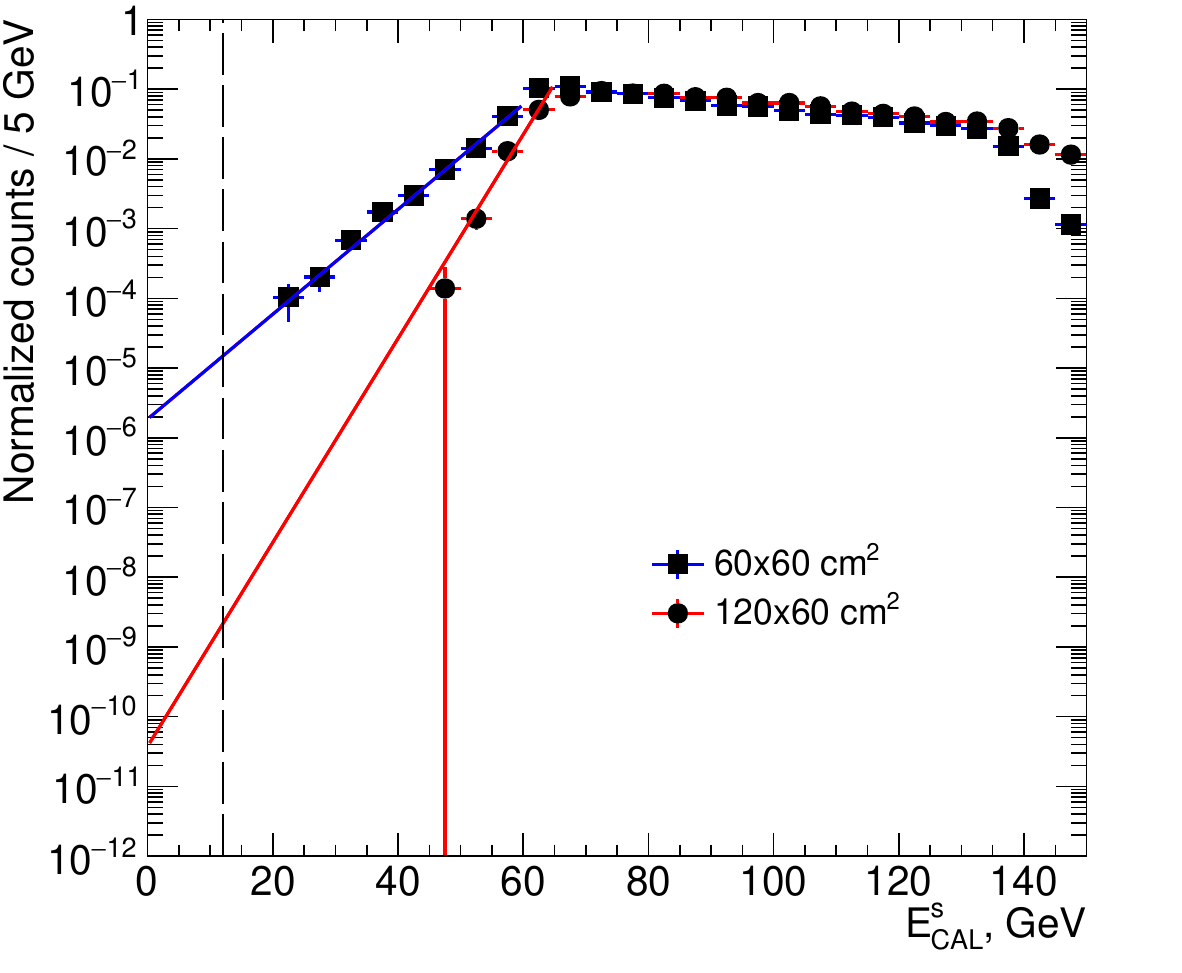}
    \caption{Energy distributions for both HCAL and ECAL from secondary products from hard muon bremsstrahlung and nuclear interactions in the target. A final-state muon with energy less than $E_\mu^\prime\leq100$ GeV is required. Both HCAL transverse sizes with 60$\times$60 cm$^2$ (blue) and $120\times60$ cm$^2$ (red) are shown. The upper bound of the signal box (black dashed line) is shown (see Sec. \ref{sec:data-analysis}).}
    \label{fig:hcal-transverse}
\end{figure}

As mentioned in Sec. \ref{sec:search-phenomenology}, the method of search relies on a well-defined final-state muon with about less than half of its initial energy, and energy compatible with that of a MIP in the sub-detectors. To maximize the hermeticity of the set-up, and avoid background due to energy leakage (see discussion of Sec. \ref{sec:background-sources}), the detector geometry has been optimized through MC simulation. In particular, the HCAL acceptance has been maximized to suppress non-hermeticity from events with hard muon bremsstrahlung or nuclear interactions in the target, with a low energetic final-state muon accompanied by large-angle emitted secondaries. This is achieved by isolating from simulation both charged and neutral secondaries produced in the target and respectively deflected through MS2 outside of the HCAL acceptance or emitted at large angles for different HCAL transverse sizes. In Fig. \ref{fig:hcal-transverse}, the energy distributions of the sum of ECAL+HCAL from interactions in the target are shown for both the NA64$e$ HCAL modules \cite{NA64:2023wbi} with longitudinal depth $7.5\lambda_\text{int}$ and lateral size $60\times60$ cm$^2$ and the newly-designed $120\times60$ cm$^2$ modules. The spectra assume a single scattered muon with energy $E_\mu^\prime<100$ GeV, and no energy deposit on the VHCAL. The reduction in non-hermeticity due to lateral energy leakage is estimated through a power law fit of the exponential tail of the distributions, showing a gain of $\sim\mathcal{O}(10^{4})$ in hermeticity with respect to the NA64$e$ geometry.
To further improve the set-up hermeticity, the VHCAL prototype was optimized through simulation to maximize the acceptance for the passage of non-interacting beam muons in the target and minimize events from upstream interactions. The $120\times60$ mm$^2$ central hole size, lateral $50\times50$ cm$^2$ dimension, and $5\lambda_\text{int}$ were extracted from MC simulation based on the ratio of beam muons to scattered secondaries. Its placements within the set-up was defined from simulations to provide nearly full hermetic coverage for secondary particles produced in the ECAL.
\section{Data analysis\label{sec:data-analysis}}
In this work, the analysis of the search for an invisible decaying $Z'$ boson from the reaction $\mu N\rightarrow\mu NZ'$ is performed on the full data sample collected during the NA64 May 2022 run. This corresponds to a total statistics of $N_\text{MOT}=(1.98\pm0.02)\times10^{10}$ gathered at a moderate beam intensity of $2.8\times10^{6}$ $\mu/$spill. In particular, the presented results are obtained by the combination of two sets of data, with different trigger configurations (hereafter called trigger 1 and 2 configurations) given the S$_\mu$ counter shifted respectively 152 mm and 117 mm along the deflection axis. This corresponds to $N_\text{MOT}^{(1)}=(1.17\pm0.01)\times10^{10}$ and $N_\text{MOT}^{(2)}=(0.81\pm0.01)\times10^{10}$, with a relative trigger efficiency of 0.04\% and 0.07\% that of the calibration trigger coincidence. More details on the analysis can be found in \cite{Sieber:2024phd}.\\ \indent
The analysis aims at identifying single final-state muons from the reaction \eqref{eq:production}, with an activity in the sub-detectors comparable with that of a MIP. This is achieved by precisely reconstructing both the incoming and outgoing momenta of such events while rejecting multi-track events, and appropriately defining MIP-compatible selection cuts to veto possible hard muon breamsstrahlung or nuclear upstream interactions. The approach is thus chosen to be that of a cut-flow-based analysis, with a blinded signal region in the kinematic variable space associated with $(p_\text{out},\ E_\text{CAL})$, with $E_\text{CAL}=E_\text{ECAL}+E_\text{VHCAL}+E_\text{HCAL}$ the total energy deposited on the calorimeters. 
\begin{table}[H]
    \centering
    \begin{tabular}{ccccccc}
    \hline
    \hline
         &   \multicolumn{5}{c}{$\kappa_{Z'}$ (\%) for $m_{Z'}$ in MeV} & \\
         \hline
        $p_\text{out}$ (GeV) & $1$ & $10$ & $10^{2}$ & $5\times10^{2}$ & $10^{3}$ & $B$ \\
        \hline
        \hline
        60 & 4.4 & 9.4 & 41.1 & 70.0 & 75.1 & $5\cdot10^{-5}$ \\
        70 & 5.7 & 12.4 & 50.1 & 77.2 & 81.9 & 0.001 \\
        80 & 7.6 & 15.9 & 58.6 & 82.2 & 86.0 & 0.05  \\
        90 & 9.7 & 19.5 & 67.1 & 86.3 & 90.1 & 1.6 \\
        100 & 12.1 & 24.8 & 74.9 & 89.7 & 93.2 & 53.0  \\
        \hline
        \hline
    \end{tabular}
    \caption{Signal efficiency $\kappa_{Z'}$ for different $Z'$ masses, and background level $B$ as a function of the choice of bound on $p_\text{out}$. The background is extracted from data (see Sec. \ref{sec:background-sources}) and the signal efficiencies from MC. The events are extracted assuming calibration trigger configuration and MIP-compatible behaviour.}
    \label{tab:sensitivity-table}
\end{table}
The signal region is defined by minimizing the number of background event $B<1$, while maximizing the signal efficiency $\kappa_{Z’}$. The optimal bound value on $p_\text{out}$ is found to be $<80$ GeV/c, and obtained by studying the $Z'$ differential energy spectrum given through Eq. \eqref{eq:analytical-muon}, and simulating the trigger acceptance as shown in Sec. \ref{subsec:beam-optics}. The background yield is studied through extrapolation from calibration trigger muon data (see Sec. \ref{sec:background-sources}). Results for different choices of $p_\text{out}$ are shown in Table \ref{tab:sensitivity-table}.The bound on $E_\text{CAL}$ is obtained by isolating and summing the muon MIP contributions in each of the sub-detectors contributing to the total energy deposit in the calorimeters and found to be $E_\text{CAL, box}<12$ GeV.\\ \indent
Appropriate selection criteria are imposed on the data sample, to both minimize the expected background sources and maximize the likelihood of observing the expected signal signature defined in Sec. \ref{sec:search-phenomenology}. Having blinded the signal region, no bias is introduced towards the search of $Z'\rightarrow\text{invisible}$. The main cuts are defined as follows,
\begin{itemize}
    \item[(i)] There must be one and only one incoming muon track, with momentum falling within the window $p_\text{in}\in[140,\ 180]$ GeV/c. Additionally, appropriate quality cuts are applied on the reconstructed tracks' underlying $p-$value distribution. This cut forces the selection before the interaction point of well-defined primary muons from the core of the beam (see Sec. \ref{subsec:beam-optics}).
    \item[(ii)] One and only one scattered muon track should be reconstructed in the second magnet spectrometer ($p_\text{out}$ in MS2), with at most a single hit in the tracking detectors MM$_{5-7}$ and ST$_{1}$ (multiplicity$\leq$1). Additionally, the tracklet associated with those trackers is extrapolated to the HCAL's face and it is verified that the energy deposit in the cell is compatible with that of a MIP. This criterion enforces the selection of a single outgoing muon with no energetic enough secondaries from upstream ECAL interactions.
    \item[(iii)] The energy deposit in both the electromagnetic and hadronic calorimeters should be compatible with that of a MIP. In addition, there should be no activity in the VHCAL, and no energy deposit in VETO different than that of a MIP.
\end{itemize}
For proper identification of incoming and outgoing tracks, it is further required that all hits are in-time, falling within the individual tracker time resolutions defined in Table \ref{tab:detectors_parameters}, with an in-time requirement of a $\delta_t\sim25$ ns time window. The distribution of events from $\mu N\rightarrow\mu+\text{anything}$ is shown in Fig. \ref{fig:hermeticity-region} \cite{Andreev:2024sgn} for both physical trigger configurations 1 and 2, for one and only one muon traversing the whole set-up (selection criteria (i+ii)) and for the whole set of cuts (i-iii). Four distinct regions are highlighted in the hermeticity plane $(p_\text{out},\ E_\text{CAL})$.
\begin{widetext}
\par\smallskip\noindent
\centerline{\begin{minipage}{\linewidth} 
\begin{figure}[H]
    \centering
    \includegraphics[width=0.37\textwidth]{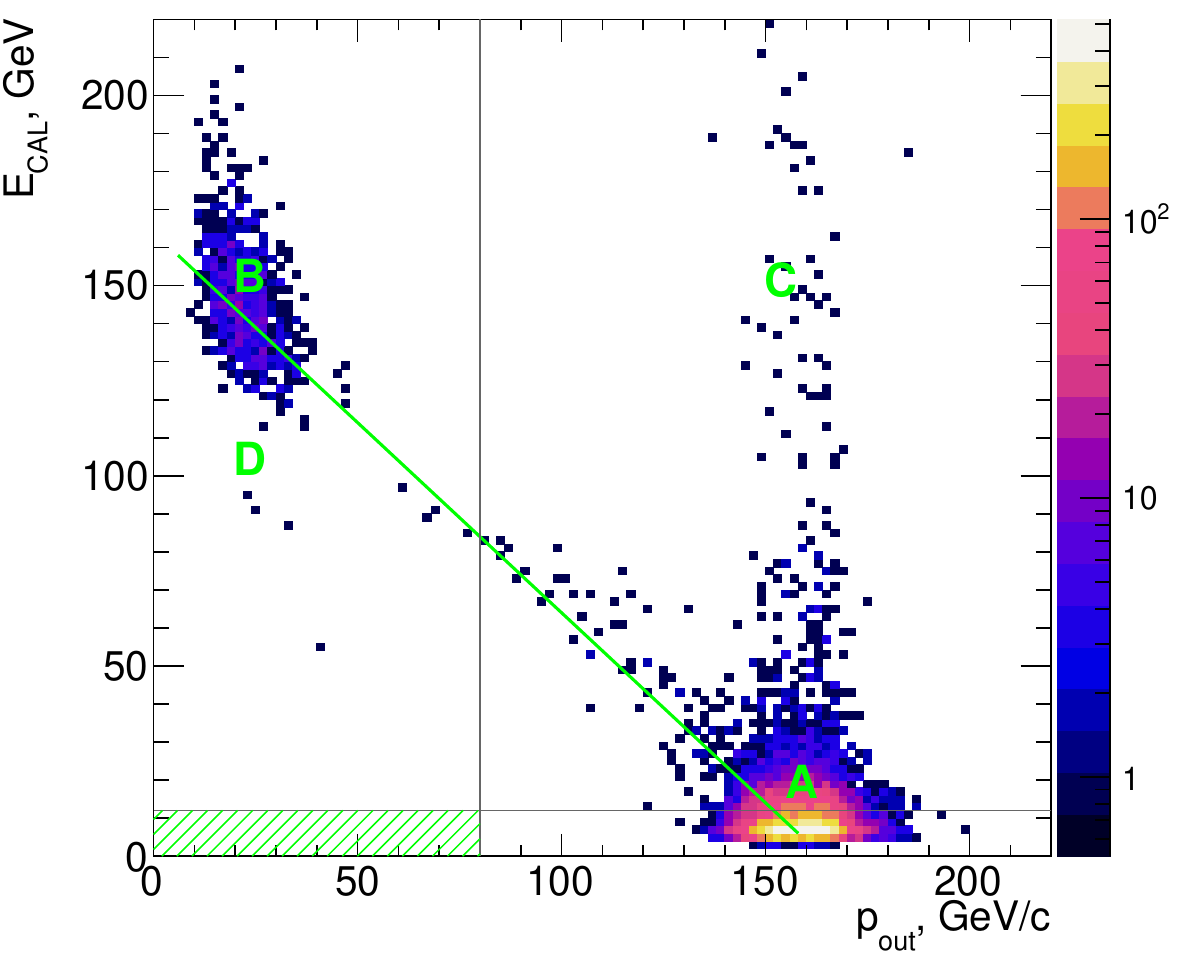}
    \hspace{1cm}
    \includegraphics[width=0.37\textwidth]{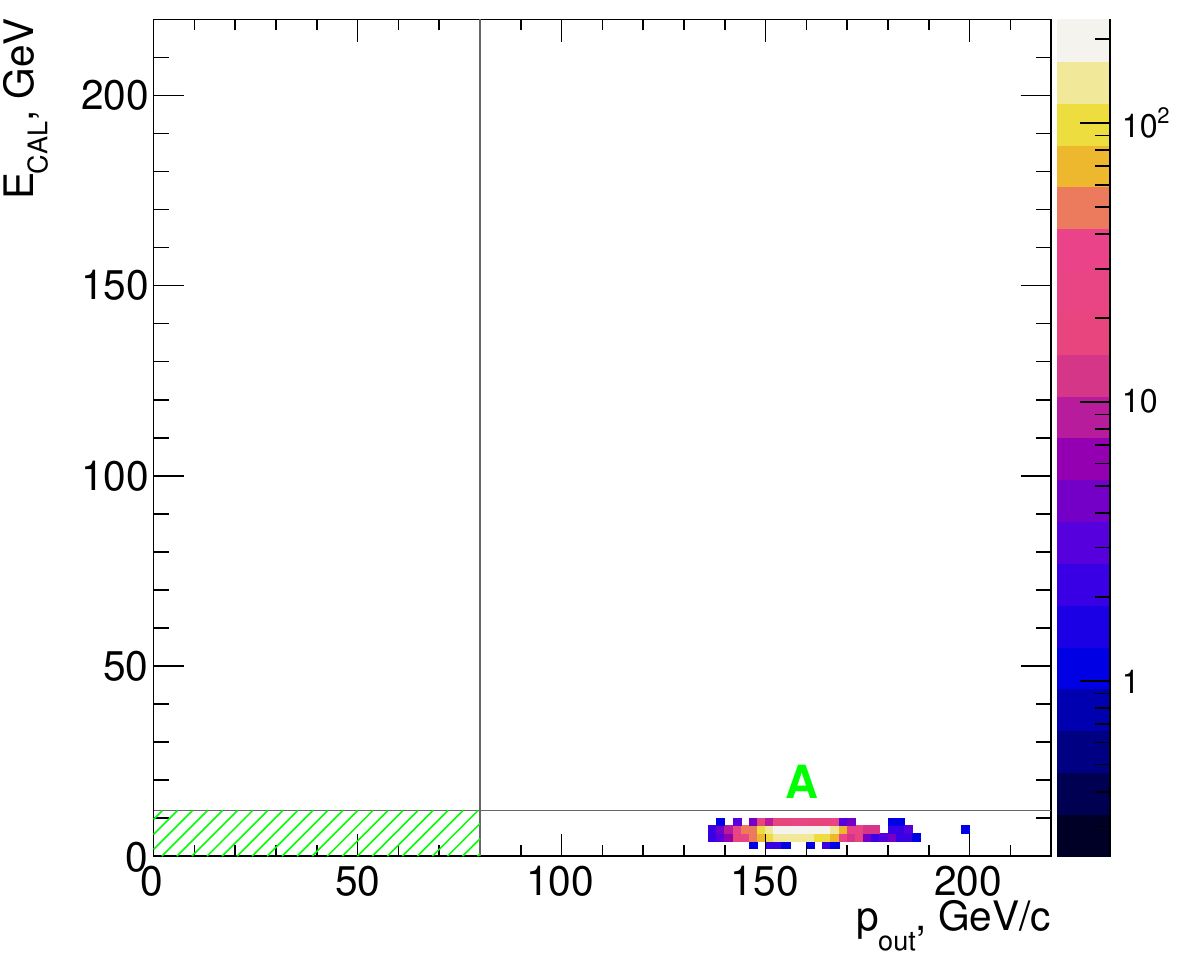}
    \caption{Event distribution in the hermeticity plane defined by the reconstructed momentum after the ECAL target, $p_\text{out}$, and the sum of energy deposit in the calorimeters, $E_\text{CAL}=E_\text{ECAL}+E_\text{VHCAL}+E_\text{HCAL}$ \cite{Andreev:2024sgn}. The sample of events corresponds to the sum of both physical trigger configurations 1+2, with $N_\text{MOT}=(1.98\pm0.02)\times10^{10}$. The signal region is blinded (green right-hashed box). Left: Events distribution after applying selection criteria (i-ii) to select a single-track-compatible event with a muon traversing the whole set-up. For completeness, different regions of the phase space are highlighted, with regions $A$ and $B$ used as \emph{control} regions for background extrapolation (see text). Right: Event distribution after additionally requiring a MIP in the calorimeters and no activity in the VHCAL and VETO (criteria (i-iii)).}
    \label{fig:hermeticity-region}
\end{figure}
\end{minipage}}
\end{widetext}
Region $A$ is inherent to MIP-compatible events traversing the whole set-up while being nearly undeflected within MS2. This relates to the condition $p_\text{in}\simeq p_\text{out}\simeq160$ GeV/c, with interactions in the ECAL and the detectors downstream of the target compatible with that of a MIP. While by design, most of those events do not pass the physical trigger condition defined by \eqref{eq:trigger-physical}, sufficiently energetic residual electrons, $\mu N\rightarrow\mu N+\delta e$, from interactions in MM$_7$, the ST$_1$ planes, or the last HCAL layer, do clinch a trigger coincidence with S$_{4}$ and S$_{\mu}$. Region $B$ corresponds to hard scattering/bremsstrahlung events, $\mu N\rightarrow\mu N+X$, with a soft muon in the final state, and a large energy deposit either in the ECAL target of the HCAL modules. Because of the shift of the trigger counters along the deflection axis past MS2, events with hard bremsstrahlung, $\mu N\rightarrow\mu N+\gamma$, and small final-state muon scattering angle, $\psi_\mu^\prime\ll10^{-2}$ rad, do not pass the physical trigger, thus resulting in a small number of events populating the momentum range $p_\text{out}\in[50,\ 100]$ GeV/c. In the event of a quasi-full energy deposit of the final-state muon solely in the HCAL modules through muon nuclear interactions, events accumulate along the horizontal axis from region $A$ to $C$. Finally, because of the limited detector acceptance of the HCAL modules, events with muon nuclear interactions in the ECAL, $\mu N\rightarrow\mu N+X$, populate region $D$. For such events, $X$ is typically any combination of mesons or baryons such as $\pi$ and $K$, or protons and neutrons ($p$ and $n$), accompanied by a low-momentum final-state muon and low-energetic charged hadrons being deflected away in MS2, thus missing the HCAL modules.
\section{Monte Carlo Validation}
The data analysis strongly relies on the accuracy of the MC simulation, in particular in the estimate of the signal yield and efficiency (see Secs. \ref{subsec:signal-production} and \ref{sec:signal-yield}). To assert the reliability of the simulation framework presented in Sec. \ref{sec:mc-approach}, and evaluate the systematics associated with the signal, the MC is benchmarked against data. As the reaction Eq. \eqref{eq:production} is associated with a single deflected muon in the final-state, both the track propagation through the magnetic fields and the MIP signatures in the calorimeters are compared.
\subsection{Track deflection}
The reconstruction of the muon initial- and final-state momenta is performed within the following constraints: (i) an 80-meter-long magnet spectrometer (MS1) with (ii) complex non-uniform magnetic fields due to the multiple quadrupoles, (iii) a large beam spread in space due to the FODO configuration of the upstream part of the experiment and (iv) a non-trivial hit multiplicity due to upstream interactions with the beam material and halo muons. The event reconstruction pipeline consists of dedicated digitization of each of the tracking detectors' hits, followed by a cellular-automaton-based track finding algorithm (CAT) implemented following the work of \cite{Abt:2002he} to cope with constraints (iii-iv). The resulting track candidates are weighted (pre-fitted) based on a singular value decomposition (SVD) scheme and fitted using a deterministic annealing filter (DAF) implemented within the \texttt{GenFit} package \cite{Rauch:2014wta}. The \texttt{OPERA} \cite{Dassault:2023mnl} output of the individual magnetic field maps (ii) is embedded within the track reconstruction sequence through a fourth-order finite difference method for field value interpolation in space. A complete overview of the track reconstruction scheme can be found in \cite{Sieber:2024phd} and a snapshot of such events shown in Fig. \ref{fig:genfit-rendering}.\\ \indent 
\begin{widetext}
\par\smallskip\noindent
\centerline{\begin{minipage}{\linewidth} 
\begin{figure}[H]
    \centering
    \includegraphics[width=0.99\textwidth]{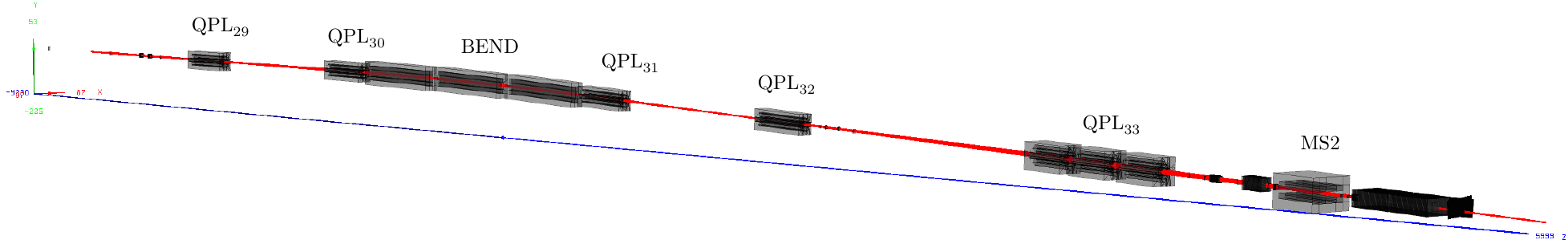}
    \caption{\texttt{GenFit}-based \cite{Rauch:2014wta} track reconstruction through the whole experimental set-up.}
    \label{fig:genfit-rendering}
\end{figure}
\end{minipage}}
\end{widetext}

\begin{figure}[H]
    \centering
    \includegraphics[width=0.37\textwidth]{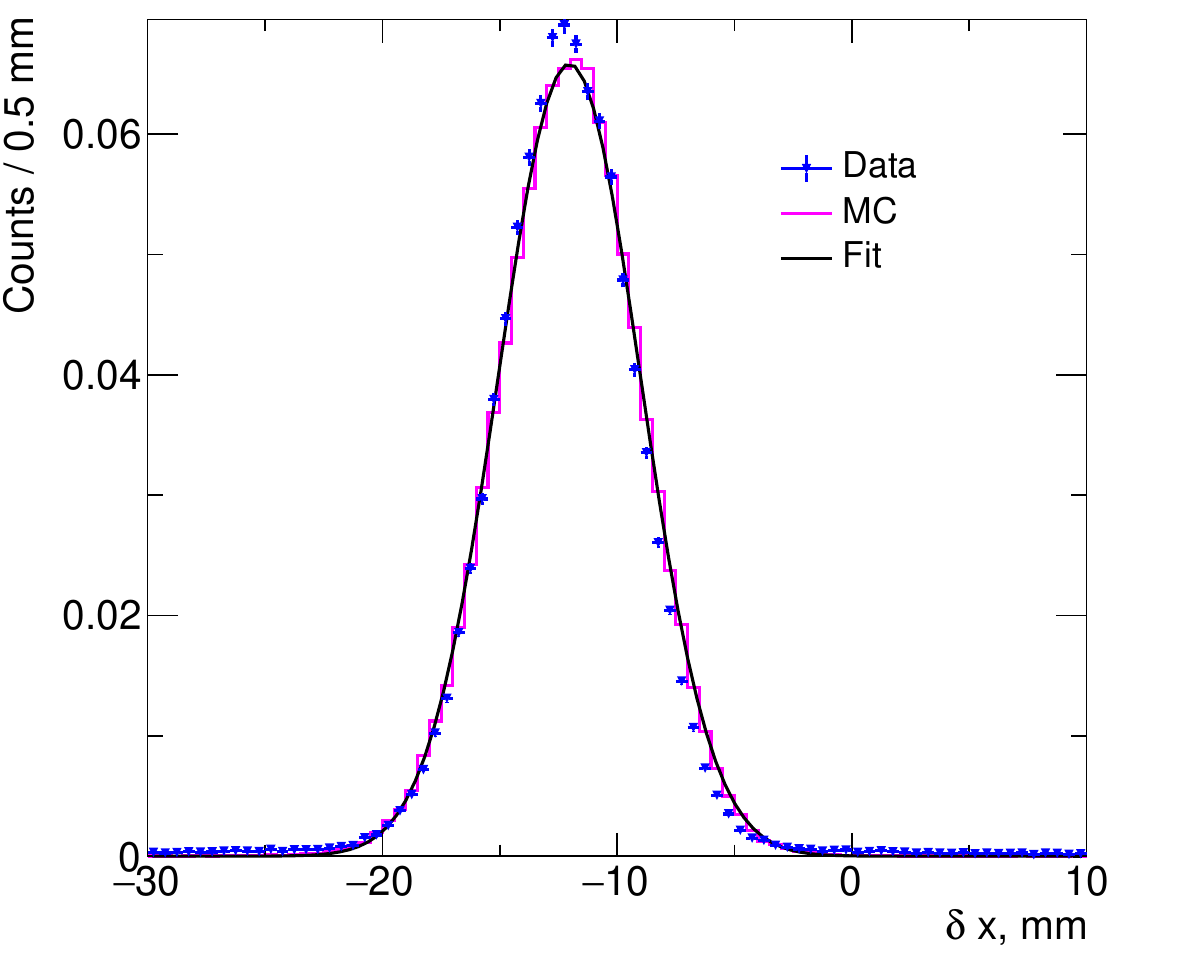}
    \caption{Distribution of single-track muon events deflection past MS2, defined as $\delta x=(\text{MM}_7)_{x}-(\text{GEM}_1)_x$, for both data (blue triangle) and MC (solid magenta line) in the calibration trigger configuration.}
    \label{fig:deflection-ms2}
\end{figure}
Using the aforementioned pipeline, the reconstructed trajectories of both MC and data events are compared in Fig. \ref{fig:deflection-ms2} to assess the effect of the fields to deflect muons, of particular importance for the estimate of the signal yield efficiency (see Sec. \ref{sec:systematics}). Samples of well-defined incoming muons satisfying selection criterion (i), with a MIP in the ECAL, are selected both from MC and data, and the typical deflection in MS2 is inferred. It is found that both spectra agree well, with a relative error $\leq2\%$ on the mean deflected position $\langle\delta x\rangle\simeq-12$ mm, between MM$_{7}$ and GEM$_1$.

\subsection{The ECAL and HCAL energy spectra}\label{subsec:ecal-hcal-spectra}
\begin{figure}[H]
    \centering
    \includegraphics[width=0.4\textwidth]{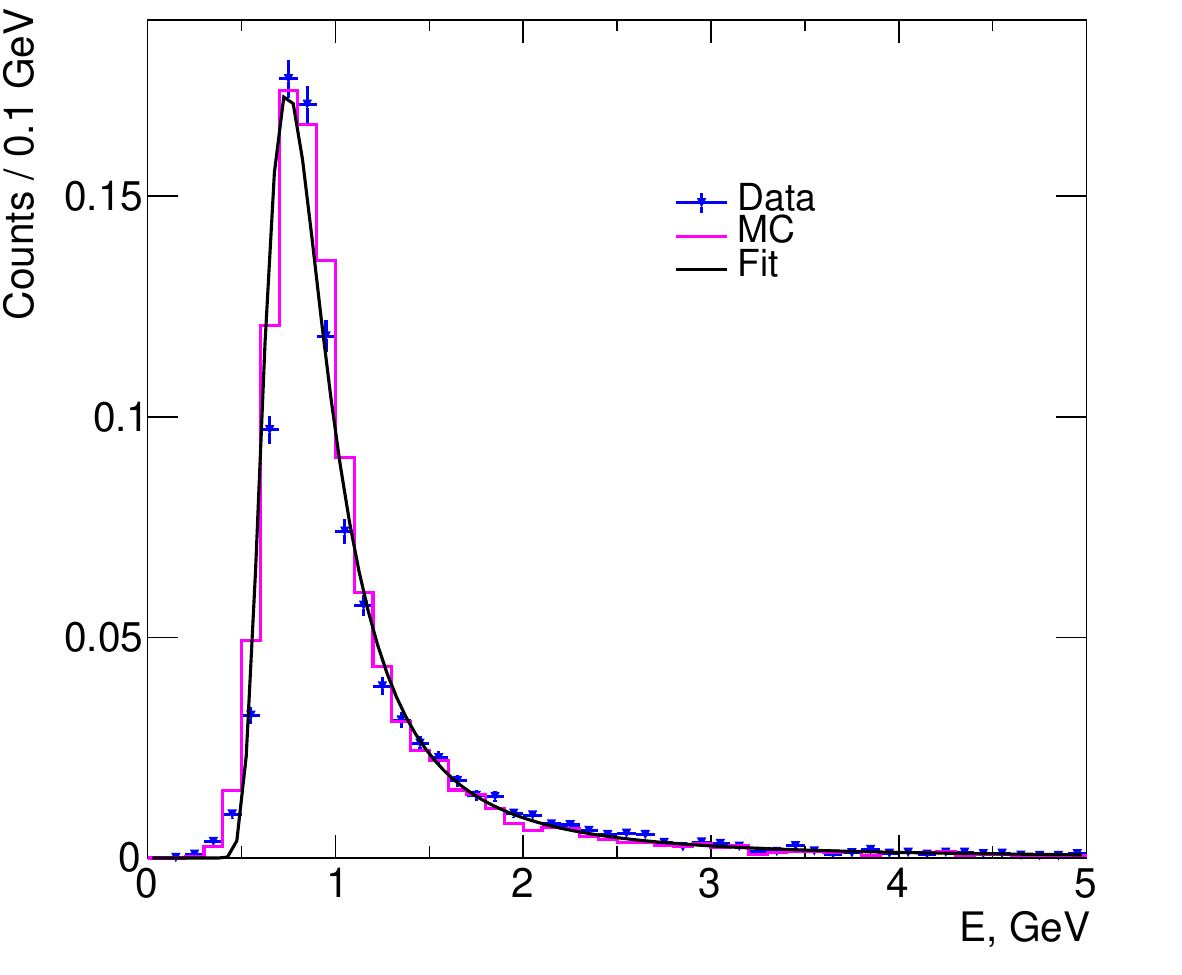}
    \hspace{1cm}
    \includegraphics[width=0.4\textwidth]{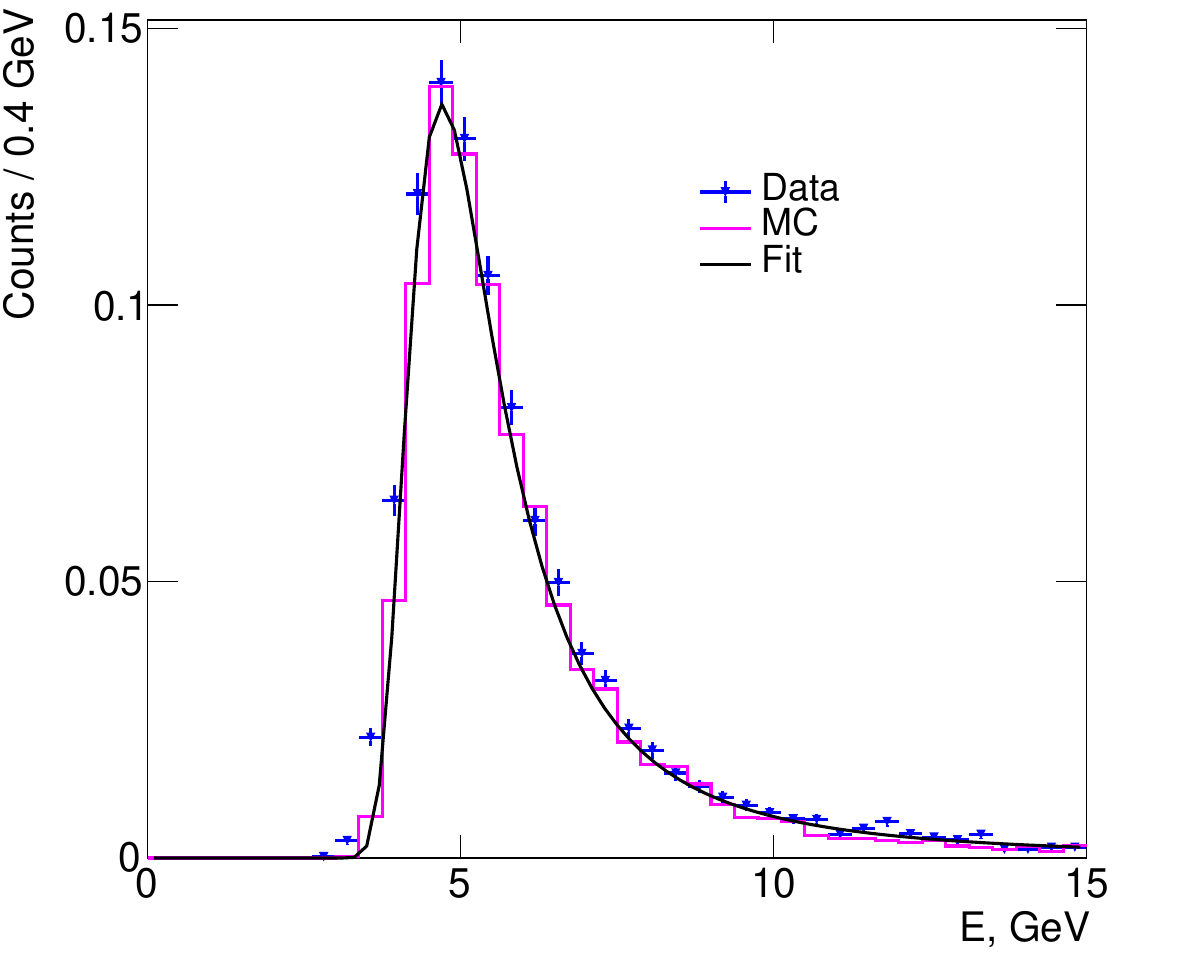}
    \caption{Distributions of energy deposited around the MIP peak compatible with a muon for both data (blue triangle) and MC (solid magenta line) in the calibration trigger configuration. Left: The ECAL module. Right: The whole HCAL module (first and second HCAL modules). The spectra are normalized to a similar number of events.}
    \label{fig:ecal-hcal-mc-data}
\end{figure}

\begin{figure}[H]
    \centering
    \includegraphics[width=0.4\textwidth]{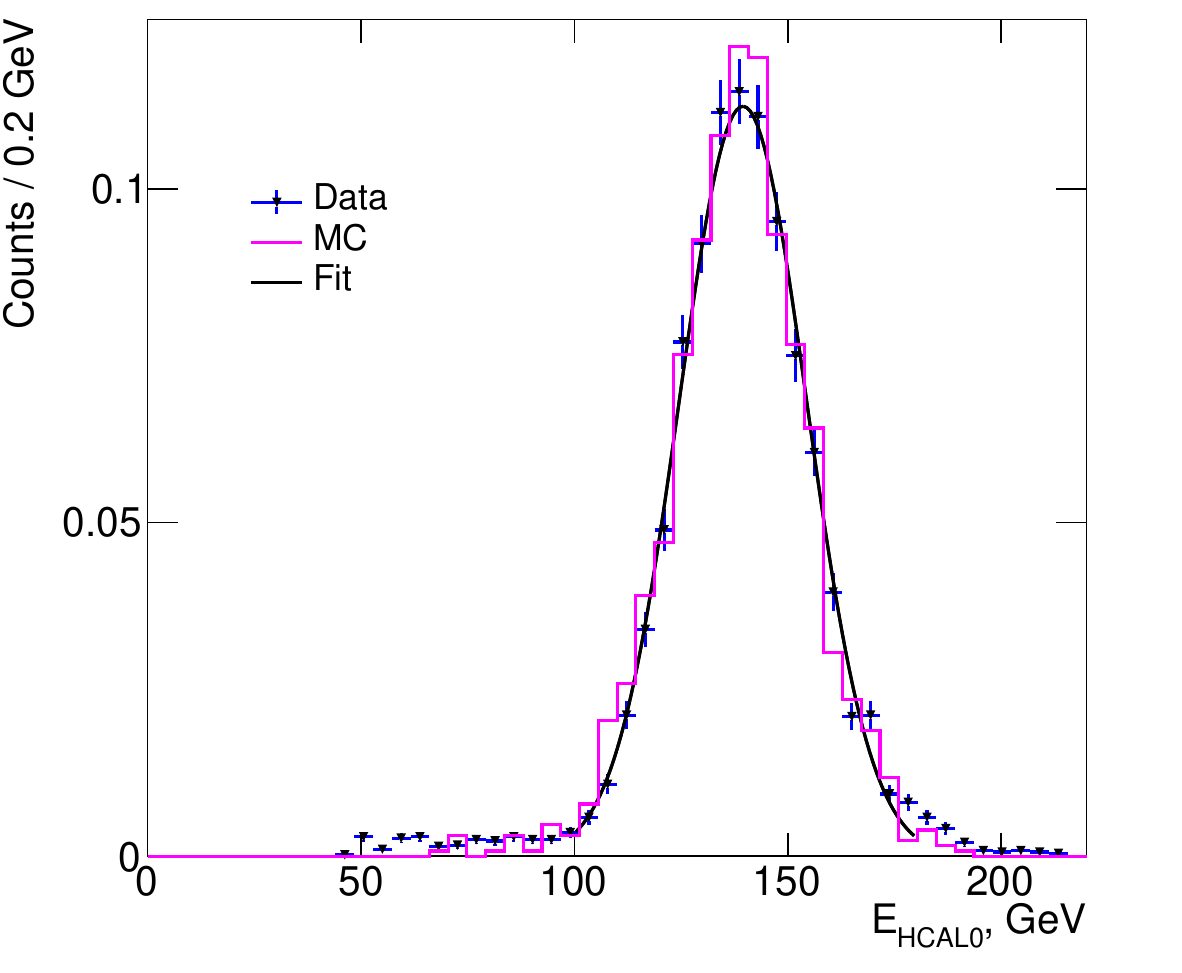}
    \caption{First HCAL module energy distribution for a sample of muon interacting in the ECAL, $\mu N\rightarrow\mu+X$, with high energetic secondaries escaping the detector and propagating through MS2. Both data (blue triangle) and MC (solid magenta line) correspond to the physics trigger configuration 1. The spectra are normalized to a similar number of events. See text for more details.}
    \label{fig:hcal0-trigger1}
\end{figure}

The agreement between MC and data is also inferred in the detector in terms of energy deposit around the muon MIP peak. This value is found to be $E_\text{ECAL}^\mu\simeq0.8$ GeV and $E_\text{HCAL}^\mu\simeq2.5$ GeV respectively for a single electromagnetic and hadronic calorimeter module. As shown in Fig. \ref{fig:ecal-hcal-mc-data}, those are well reproduced by the MC with a relative error $\leq1\%$. Additionally, both MC and data distributions are integrated around the MIP peak, resulting in a ratio of integrated events $\simeq1.03$.
While the behavior of single muons energy loss through ionization is well reproduced within the detector, the simulation is further validated using as a benchmark process hard muon nuclear interaction in the ECAL, $\mu N\rightarrow\mu+X$, where $X$ escape the target and propagate downstream of the target (see Secs. \ref{sec:data-analysis} and \ref{sec:background-sources}). $X$ is associated with either secondary hadrons, with a leading high-energetic hadron in the final state ($E_h\geq80$ GeV, see Sec. \ref{sec:background-sources}), or a highly energetic photon produced through muon-bremsstrahlung in the last layer of the ECAL (see Sec. \ref{subsec:beam-optics}). In both scenarios, the final-state muon has low momentum. A sample from both data and MC in physics trigger configuration 1 is selected, requiring (i) a muon having a momentum past ECAL $p_\text{out}\leq80$ GeV/c, (ii) no activity in the VHCAL, (iii) an energy deposit in the HCAL $E_\text{HCAL}\geq50$ GeV, and (iv) most of the energy being absorbed in the first HCAL module (HCAL$_{0}$). The resulting HCAL$_{0}$ distributions are shown in Fig. \ref{fig:hcal0-trigger1}. While the bulk of the spectra are in good agreement, small discrepancies between the data and MC appear in the tails. Those are dominated by the alignment of the trigger counters S$_{4}$ and S$_{\mu}$ (see Fig. \ref{fig:trigger-physical} and Sec. \ref{sec:systematics}).
\section{Background\label{sec:background-sources}}
The background sources associated with the search for $Z'\rightarrow\text{invisible}$ require a careful treatment to identify the single muon trajectory through the set-up. The level at which those events are expected is estimated per MOT from both detailed MC simulations and data. The MC study of the background level is based on our previous work \cite{Sieber:2021fue}.
\begin{itemize}
    \itemsep0em
    \item[(i)] \emph{Single-hadron punch-through}. Missing energy events can originate from highly energetic \emph{leading} hadrons, $h$, produced in the ECAL target through muon nuclear interaction, $\mu N\rightarrow\mu+X+h$. In such a background process, the leading hadron carries away a significant fraction of the muon energy, $E_h\geq80$ GeV, and leaks through the detectors downstream of the target with two possible scenarios: (i) shall $h$ be charged, it can be accompanied, while depositing a MIP-compatible energy in the HCALs, by a low energetic outgoing muon being poorly detected or (ii) shall $h$ be neutral, it can traverse the HCAL modules undetected while the outgoing muon is reconstructed with low energy ($p_\text{out}\leq80$ GeV/c). This background is evaluated by combining both MC and data. The probability to produce a leading hadron, $P_h$, is extracted from simulations of hard muon interactions in the ECAL and is found to be respectively $P_h^{0}=(6.90\pm0.11)\times10^{-7}$ and $P_{h}^{\pm}=(2.61\pm0.07)\times10^{-7}$ per MOT for neutral and charged hadrons, such that $P_h\simeq10^{-6}$ conservatively. Similarly, the probability to punch-through a single HCAL module with $7.5\lambda_\text{int}$, $P_\text{pt}^{(1)}$ is estimated as $\lesssim10^{-3}$ per MOT, while for two modules $P_\text{pt}^{(2)}\lesssim10^{-6}$. Those results are compared with the punch-through probability computed from the available data \cite{Denisov:1973zv,Carroll:1978hc} on the measurement of the hadronic absorption cross-sections $\sigma_a$, such that $P_{pt}=\exp{-d/\lambda_a}$, with $d$ the distance within the HCAL module, and $\lambda_a$ the absorption length. The overall probability to observe a background event $n_\text{bkg}^{h}$ is thus estimated as
        \begin{equation}
        \label{eq:kaons-decay-bkg}
        n_\text{bkg}^{h}=\sum_{t=1,2}P_{h}\times P_\text{pt}^{(2)}\times\kappa_{\text{S}_0\text{S}_1}\times N_\text{MOT}^{t},
    \end{equation}
    with $\kappa_{\text{S}_0\text{S}_1}\simeq0.55$ the efficiency of track reconstruction extracted from data. For the full sample of events, it is found that $n_\text{bkg}^{h}=(2.8\pm0.1)\times10^{-3}$.
    \item[(ii)] \emph{Dimuons production}. Besides muon nuclear interactions in the target, potential background from muon electromagnetic processes in the ECAL can contribute to mimicking signal events, especially in the visible search for $Z'\rightarrow\mu^{+}\mu^{-}$. Dileptons events with a muon pair in the final state are associated with dimuons production through (i) the emission of a real photon (Bethe-Heitler mechanism \cite{Bogdanov:2006kr,Kelner:1997cy}), $\mu N\rightarrow\mu N+\gamma;\gamma\rightarrow\mu^{+}\mu^{-}$, (ii) the conversion of a virtual photon (Trident process \cite{Kelner:1995hu}), $\gamma^{\ast}\rightarrow\mu^{+}\mu^{-}$ and (iii) sufficiently energetic knock-on electrons yielding a bremsstrahlung-like dimuon production, $\delta eN\rightarrow\delta eN+\gamma;\gamma\rightarrow\mu^{+}\mu^{-}$ \cite{Chaudhuri:1965hb}. The yield for such events is obtained through simulations and found to be $\sim10^{-7}$ per MOT, suppressed by a factor $(m_e/m_\mu)^{5}$ with respect to the electron bremsstrahlung. For the set of cuts defined in Sec. \ref{sec:data-analysis}, double- and triple-MIP events in the calorimeters are efficiently rejected, as well as events with muons emitted at large angles leaving a signature in VETO and VHCAL, not passing criterium (iii). Similarly, track multiplicity strongly suppresses dimuons events, following selection criteria (ii). For the full statistics of $1.98\times10^{10}$ MOT, it is found from simulations that $n_\text{bkg}^{2\mu}\ll0.1$ after applying all selection criteria (i-iii).
    \item[(iii)] \emph{Detectors non-hermeticity}. The level of non-hermeticity, i.e. of energy leakage from the detectors due to lack of acceptance, is inferred within the plane defined by $(p_\text{out},\ E_\text{CAL})$ (see Fig. \ref{fig:hermeticity-region}) and the control regions $B$ and $A$. A sample of events from both trigger 1 and 2 configurations is chosen, under the assumption (i+ii), and selecting only MIP-compatible events within the target as to isolate events downstream of MS2 leaking within the signal box (suppression of region $D$ events). The energy distribution associated with region $B$ is then projected on the $E_\text{CAL}$ axis and the tails are extrapolated towards the upper limit of the signal region, $E_\text{CAL, box}=12$ GeV. Similarly, as in the previous point, the estimated background for the full statistics is found to be $n_\text{bkg}^\text{CAL}<0.01$. 
    \item[(iv)] \emph{Hadron in-flight decays}. Hadrons contaminating the M2 beam line and their subsequent (semi-)leptonic in-flight decays to final states with muons, $h\rightarrow\mu+X$, contribute to events with large missing energy. In the case of decays within the region spanning from the end of the MS1 magnet and the start of the ECAL target, such events can mimic the signature \eqref{eq:production}, in particular for $X$ being associated with neutrinos, carrying a large fraction of the available energy of the hadron, $p_\text{in}$. Since the dominant hadronic component of the beam is associated with $\pi$ and $K$, the muon spectra from beam hadron decays, $h\rightarrow\mu+X$, with $h=\pi,K$, is simulated within the framework described in Sec. \ref{sec:mc-approach}. The corresponding muon energy distributions are shown in Fig. \ref{fig:hadron-decays}, for which it is found that muons from $\pi\rightarrow\mu+X$ do not enter the final-state momentum upper limit of the signal box defined in Sec. \ref{sec:data-analysis}. The probability for charged kaons to decay, $K\rightarrow\mu+\nu_\mu$, before the ECAL target is estimated analytically through Eq. \eqref{eq:kaon-decay}, such that 
    \begin{equation}
        \label{eq:kaon-decay}
        \begin{split}
            &P(d;\ K\rightarrow\mu+\nu_\mu)=\text{Br}(K\rightarrow\mu+\nu_\mu)\\
            &\times\int_{E_\text{min}}^{E_\text{max}}dE\ f(E;\mu,\sigma)\bigg[1-\exp\bigg(-\frac{d}{\gamma\beta c\tau_{K}}\bigg)\bigg],
        \end{split}
    \end{equation}
    with $\text{Br}(K\rightarrow\mu+\nu_\mu)\simeq0.64$ \cite{ParticleDataGroup:2022pth}, $E_\text{min}$ and $E_\text{max}$ the energy bounds of the kaons, $f(E;\mu,\sigma)$ the underlying kaon energy PDF with $\mu=160$ GeV/c, $\sigma=3$ GeV/c, $l_\tau=\gamma\beta c\tau_{K}$ the proper decay length, and $d=20$ m. As such, given the hadron contamination $P_c$ and kaons-to-pions ratio, $P_{K/\pi}$, discussed in Sec. \ref{sec:na64mu-detectors}, the expected number of decays per MOT is $P=P_c\times P_{K/\pi}\times\text{Br}_{\mu+\nu_\mu}\times P_{K\rightarrow X}\simeq1.6\times10^{-8}$, with $P_{K\rightarrow X}\simeq0.017$. Integrating the $K\rightarrow\mu+X$ spectrum from Fig. \ref{fig:hadron-decays} up to $p_\text{out, box}=80$ GeV/c, and isolating the neutrinos component, this probability reduces to $5.6\times10^{-9}$ per MOT. The final background estimate is further assessed through a detailed MC study of the signature of such final-state decay muons in the experimental setup. In addition to the analytical estimate, this in particular accounts for both the beam divergence and decay angle at the level of ECAL, as well as the effects of applying the full set of cuts discussed in Sec. \ref{sec:data-analysis}. Considering both trigger configurations $t=1,2$, and the track reconstruction efficiency extracted from data, the expected number of events results in  $(8.7\pm0.7)\times10^{-3}$ for the full statistics.
    \begin{figure}[H]
        \centering
        \includegraphics[width=0.4\textwidth]{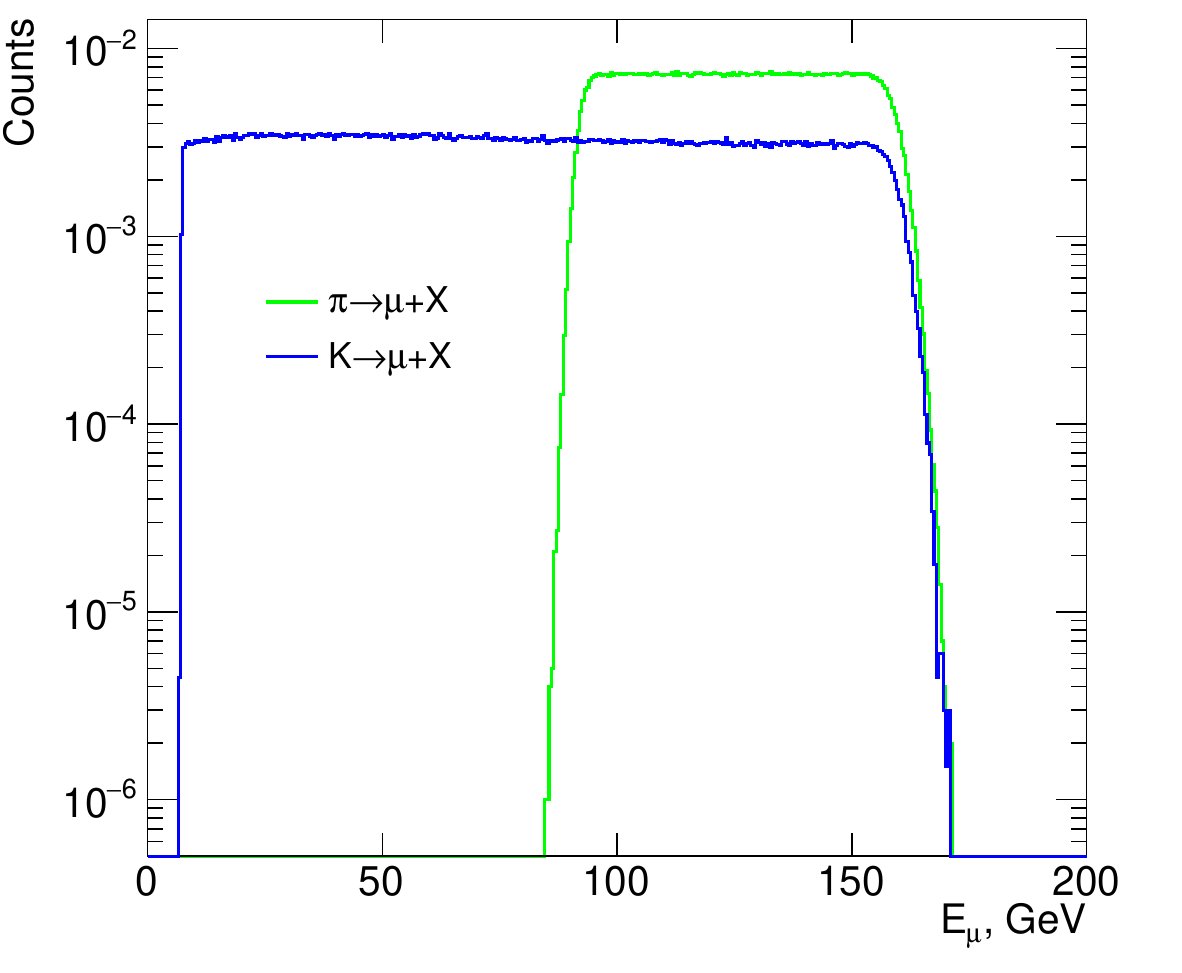}
        \caption{Final-state muon energy distributions for both pions (green line) and kaons (blue line) decays, $\pi,K\rightarrow\mu+X$, as extracted from a dedicated \texttt{GEANT4} simulation of the process within the experimental set-up.}
        \label{fig:hadron-decays}
    \end{figure}
    \item[(v)] \emph{Momentum mis-reconstruction}. The background originating from fake low-momentum tracks is associated with the misidentification of the scattered muon after the ECAL. As such, it is associated with an event with a well-defined incoming muon, $p_\text{in}\simeq160$ GeV/c, while its final-state post the interaction point is reconstructed with energy $\leq80$ GeV/c, while it truly was $\sim160$ GeV/c. This estimate is performed by selecting a sample of events, $N_\text{sample}$, within the calibration trigger configuration with selection criteria (iii) and (i), further shrinking the initial momentum window to be $p_\text{in}=160\pm10$ GeV/c . The low-energy tail of the distribution of $p_\text{out}$ corresponding to this sample is then fitted and extrapolated towards the upper limit of the signal box, $p_\text{out, box}=80$ GeV/c (see Fig. \ref{fig:momentum-extrap}), similarly as in our previous work \cite{NA64:2018iqr}. As such the background level is given by
    \begin{equation}
        \label{eq_momentum-bkg}
        n_\text{bkg}^{p_\text{out}}=\sum_{t=1,2}\frac{N_{\leq80}}{N_\text{sample}}\times N_\text{MOT}^{t},\ N_{\leq80}=\int_{0}^{80}dp\ \hat{f}(p),
    \end{equation}
    with $\hat{f}(p)$ the underlying fit function. In order to account for the systematics associated with the choice of the fit, both an exponential and a Crystal Ball functions are applied to the tails of the distribution. For the full physical statistics, the resulting background level is found to be $n_\text{bkg}^{p_\text{out}}=0.045\pm0.031\ \text{(stats)}\ \pm 0.012\ \text{(sys)}$.
    \begin{figure}[H]
        \centering
        \includegraphics[width=0.4\textwidth]{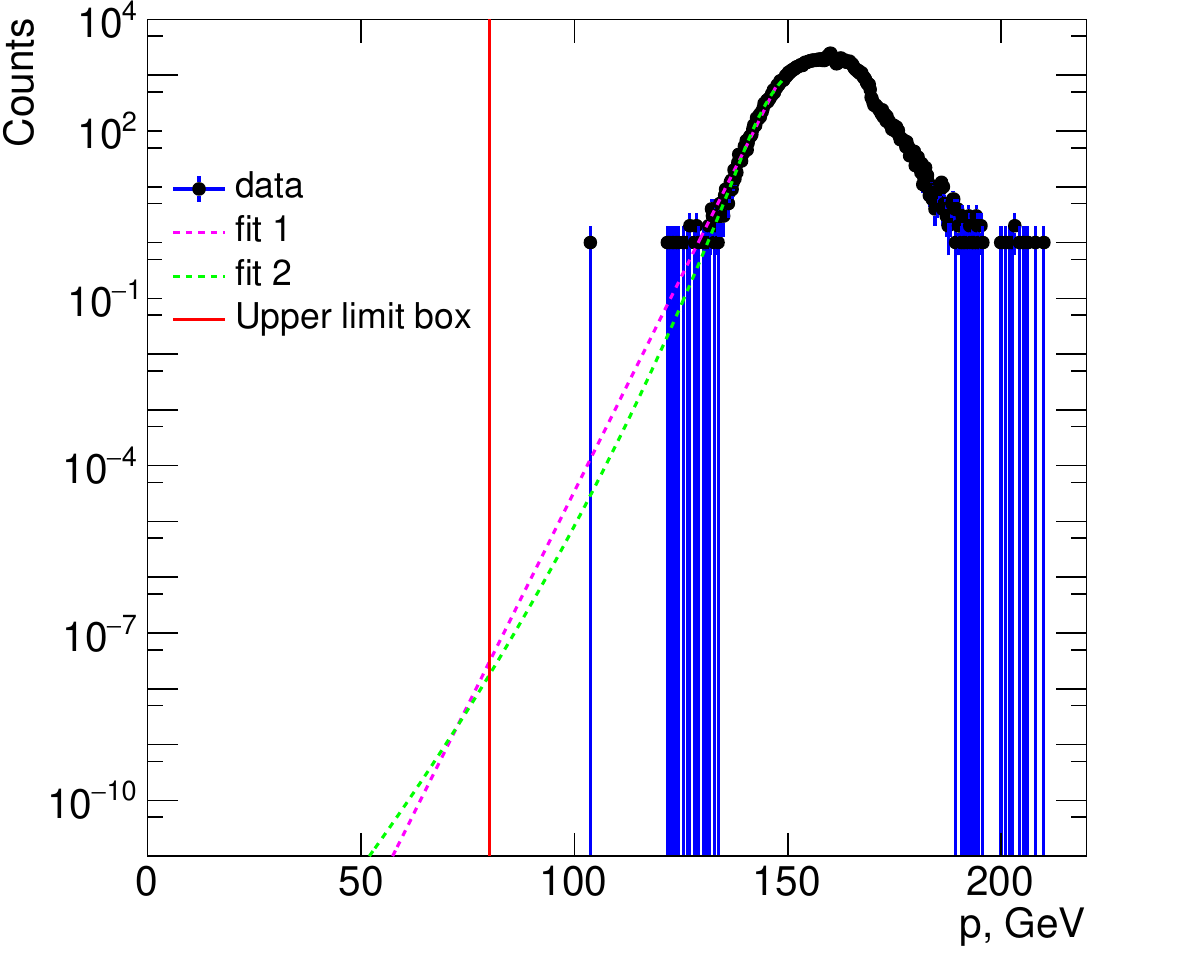}
        \caption{Exponential (dashed magenta line) and Crystal Ball (dashed green line) fits of the low-energy tails of the scattered muon momentum distribution, after applying selection criteria (iii) and (i), with $p_\text{in}=160\pm10$ GeV/c, on a sample of events extracted from the calibration trigger configuration.}
        \label{fig:momentum-extrap}
    \end{figure}
\end{itemize}
The estimated background level is given in Table \ref{table:background} as the sum of the contributions from the dominant background sources estimated using both MC and data. Adding those sources through quadrature it is found that for a total statistics of $1.98\times10^{10}$ MOT, $n_\text{bkg}=0.07\pm0.03$, with the main contribution being associated with momentum mis-reconstruction. 
\begin{table}[H]
    \centering
    \begin{tabular}{lr}
    \hline
    \hline
    Background source & Background, $n_\text{bkg}$ \\
    \hline
    (i) Single-hadron punch-through & $\ll0.01$ \\
    (ii) Dimuons production & $\ll0.01$ \\
    (iii) Detectors non-hermeticity & $<0.01$ \\
    (iv) Hadron in-flight decays & $0.010\pm0.001$ \\
    (v) Momentum mis-reconstruction & $0.045\pm0.033$ \\

    \hline
    Total (conservatively) $n_\text{bkg}$ & $0.07\pm0.03$ \\
    \hline
    \hline
    \end{tabular}
    \caption{\label{table:background}Expected background for a total of $1.98\times10^{10}$ MOT. The uncertainties are added through quadrature.}
\end{table}
\section{Signal yield\label{sec:signal-yield}}
\begin{figure}[H]
    \centering
    \includegraphics[width=0.4\textwidth]{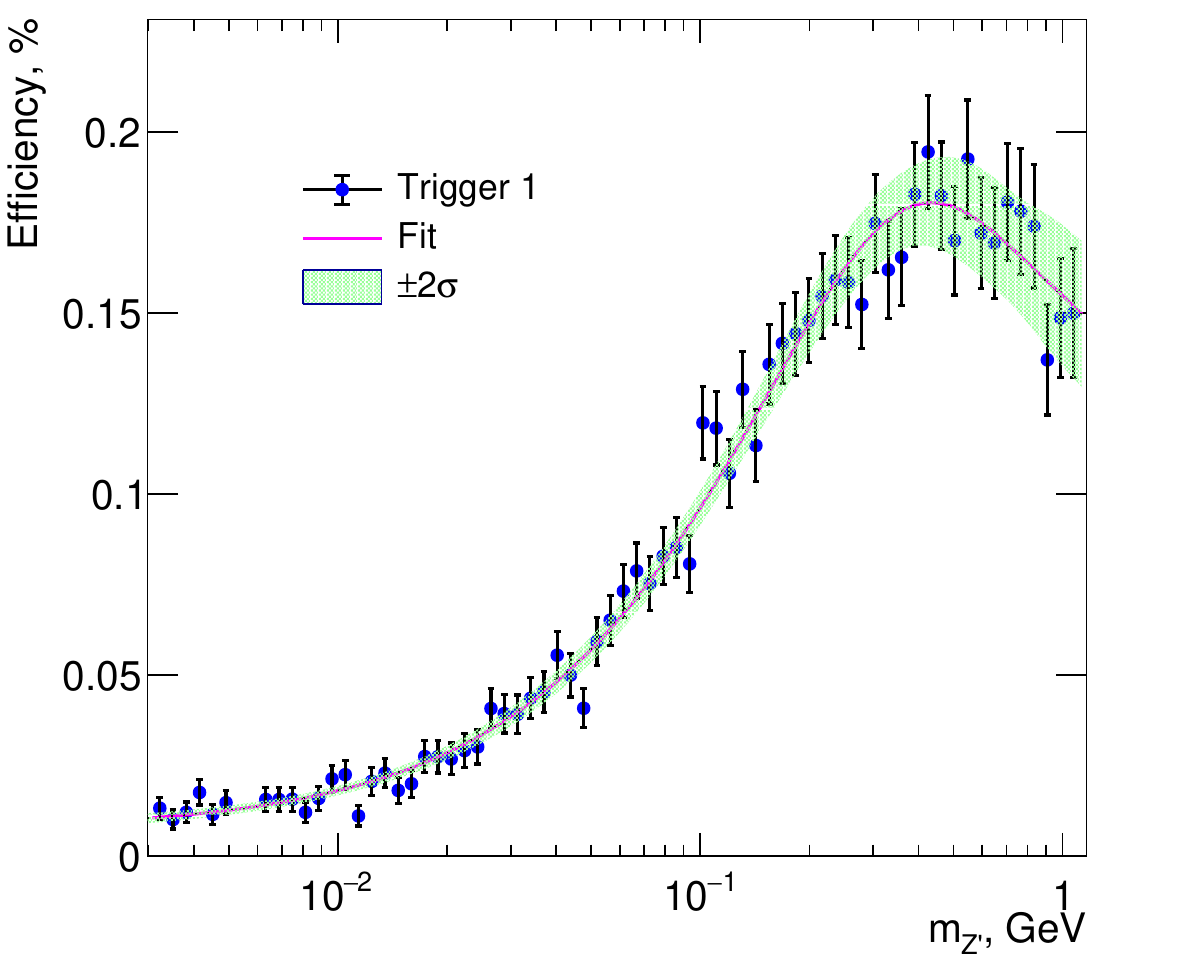}
    \caption{Effect of the geometrical acceptance associated with trigger 1 (S$_4$ and S$_\mu$ shifted respectively 65 mm and 152 mm along the magnetic deflection direction) on the signal efficiency computed as a function of the mass. The track reconstruction efficiency is also considered.}
    \label{fig:geom_signal_efficiency}
\end{figure}

The signal yield is computed by combining both data and MC such that
\begin{equation}
\label{eq:signal-yield}
    \mathcal{N}_{Z'}=N_\text{MOT}\times\kappa_{\text{S}_{1}\text{S}_{0}}\kappa_{Z'}\times\left(\frac{10^{-4}}{\epsilon_{Z'}}\right)^2\times N.
\end{equation}
The first two terms of Eq. \eqref{eq:signal-yield} are extracted from data, with $N_\text{MOT}$ being obtained by translating the number of recorded spills to equivalent MOT. The upstream efficiency, $\kappa_{\text{S}_{1}\text{S}_{0}}$, is extracted from calibration runs with trigger conditions $\text{S}_0\times\text{S}_1\times\overline{\text{V}_0}$ to properly take into account the track reconstruction efficiency before the interaction point in the ECAL, given that signal is solely generated in the downstream part of the experiment. The signal efficiency, $\kappa_{Z'}$, is extracted from individual simulations of $Z'$ mediator production and propagation within the set-up, each with different mass $m_{Z'}$ parameters to underline the dependence on the production cross-sections and muon emission angle depicted in respectively Eqs. \eqref{eq:analytical-muon} and \eqref{eq:WWypsi} (see Fig. \ref{fig:geom_signal_efficiency}). The cumulative signal efficiency after applying all selection criteria is given in Table \ref{table:signal-efficiency} for different mass points. Because of the mixing strength $\epsilon_{Z'}$ appearing as a multiplicative factor in Eq. \eqref{eq:analytical-muon}, the simulations are performed for a fixed value of $10^{-4}$. Finally, since a bias in the production cross-section is introduced to observe signal events at a reasonable rate, an overall multiplicative factor $N$ is applied to the signal yield for proper normalization. 
\begin{table}[H]
    \centering
    \begin{tabular}{lccr}
    \hline
    \hline
    $m_{Z'}$ & $\kappa_{Z'}^{(1)}$ (\%) & $\kappa_{Z'}^{(2)}$ (\%) & $N$ \\
    \hline
     10 MeV & 1.3 & 1.3 & $6.4\times10^{-10}$ \\
     100 MeV & 5.8 & 5.9 & $9.9\times10^{-11}$ \\
     500 MeV & 9.1 & 9.3 & $6.8\times10^{-12}$ \\
     1 GeV & 7.1 & 7.6 & $1.1\times10^{-12}$ \\
    \hline
    \hline
    \end{tabular}
    \caption{Cumulative efficiencies (after all cuts) for different mass points in trigger 1 and 2 configurations. The norm $N$ is also given for the benchmark $\epsilon_{Z'}$ value $10^{-4}$.}
    \label{table:signal-efficiency}
\end{table}
\section{Signal systematics\label{sec:systematics}}
The signal systematics are determined to properly normalize the signal yield associated with the bremsstrahlung-like production of $Z'$. Those are mostly related to the underlying $Z'$ production mechanism, and the difference in the detectors' response between MC and data. The main systematics are summarized in Table \ref{table:uncertainties} and reported in detail below
\begin{itemize}
    \itemsep0em
    \item[(i)] \emph{Counting MOT}. The uncertainty on the number of MOT, entering in the signal yield estimate defined in Eq. \eqref{eq:signal-yield}, is dominated by the VERSA Module Euro (VME) card scaler used to record the trigger coincidences, as the high level of purity of the M2 beam line makes the contribution of hadron and $e^{\pm}$ admixtures negligible \cite{Doble:1994np}. Because of the high accuracy of $\sim$100 MHz counting rate of the VME scaler, the probability of miscount can be neglected given a counting rate $<1$ MHz (see Sec. \ref{sec:na64mu-detectors}). On the other hand, missing information from the VME scaler can be associated with DAQ system crash and abnormal termination of the recorded run. This is conservatively estimated to be $\leq1\%$. 
    \item[(ii)] \emph{The $Z'$ physics}. The simulation of the $Z'$ production defined in Secs. \ref{sec:search-phenomenology} and \ref{subsec:signal-production} depends on the chosen WW approach of the computations. From previous works, \cite{Kirpichnikov:2021jev,Sieber:2023nkq}, the uncertainty of the WW phase space approximation can be estimated to be $\leq2\%$ with respect to the ETL.
    Additionally, systematics shifts on the ETL originate from both the running of $\alpha$ and QED higher orders corrections related to soft photon emissions. In the former case, the contribution of $\alpha(Q^2)$, with $Q^2$ some energy scale, can be extracted from the ETL vertex factors of the Feynman diagram associated with $\mu N\rightarrow\mu NZ'$. The vertex of the $Z'$ boson coupling to muons contributes as $g_{Z'}$, while the muon-nucleon vertices contribute respectively as $e$ and $eZ$, resulting in a signal yield $\mathcal{N}_{Z'}\sim\alpha^2Z^2g_{Z'}^2$. Given the running of $\alpha$
    \begin{equation}
    \alpha(Q^2)\simeq\frac{\alpha(0)}{\big(1-\frac{\alpha(0)}{3\pi}\log\frac{Q^2}{m_e^2}\big)},
    \end{equation}
    with $m_e$ the mass of the electron, $\alpha(0)=1/137$, and for an upper bound on $Q^2\simeq m_{Z'}^2\sim\mathcal{O}(1\ \text{GeV}^2)$, the associated systematic shift contributes as a positive $2.4\%$ corrections to the signal yield, $\mathcal{N}_{Z'}$. The systematics associated with higher order corrections can be extracted from \cite{Peskin:1995ev}, for which both the contribution of virtual and real photons from the scattering process $p\rightarrow p'+\sum_{n}p_\gamma$ are considered. In particular, the probability of emitting photons with energy below the photon detector threshold $\sum_{n}p_\gamma<E_\gamma$ is given by
    \begin{equation}
    \begin{split}
    &(d\sigma)_\text{meas.}=\\
    &(d\sigma)_\text{ETL}\exp\big(-\frac{\alpha}{\pi}\log\frac{-q^2}{m_e^2}\log\frac{-q^2}{E_\gamma^2}\big),\\
    \end{split}
    \end{equation}
    with $q^2=p-p'$ and $-q^2\gg m_e^2$. For $-q^2=m_{Z'}^2\sim\mathcal{O}(1\ \text{GeV}^2)$, $m_e=m_\mu$ and $E_\gamma=500$ MeV, the associated systematic shifts is given by a negative Sudakov correction of $1.4\%$. Finally, because of the dependence of the signal yield on the target material, $Z^2$, small systematics are associated with the purity of the ECAL absorber layers (Pb plates), conservatively estimated to be $\leq1\%$. As such, the total uncertainty on the signal yield associated with the $Z'$ physics is obtained through a quadratic sum of the above to be $\leq4\%$.
    \item[(iii)] \emph{Detectors' response}. Different detectors' responses between MC and data, such as energy scales, impact the signal efficiency $\kappa_{Z'}$, especially through the corresponding choice of energy threshold. The associated systematics are evaluated around the MIP-compatible peak in both the ECAL and HCAL distributions for MC and data (see Sec. \ref{subsec:ecal-hcal-spectra}), through spectra integration and peak ratio. It is found to be $\sim3\%$ for the individual calorimeter, thus yielding a cumulative systematic of $4\%$.
    \item[(iv)] \emph{Alignment effects}. The method of search for $Z'\rightarrow\text{invisible}$ strongly relies on momentum reconstruction, for which the final-state muon momentum, and thus deflection, is closely correlated with the trigger configuration, and signal efficiency $\kappa_{Z'}$. The associated systematics are estimated both by comparing the expected deflection past MS2 in MC and data and by quantifying the effects of misalignment of the trigger counters S$_{4}$ and S$_\mu$. In the former case, the uncertainty is extracted from a data sample of well-defined MIP-compatible 160 GeV/c muons following selection criteria (i-ii), passing the calibration trigger condition, and compared with MC (see Fig. \ref{fig:deflection-ms2}, where the distributions are fitted under the Gaussian hypothesis, with a mean displacement $\langle\delta x\rangle\simeq-12.1$ mm). By additionally comparing the distributions in the transverse direction to the deflection, it is found an uncertainty of $\sim1\%$ between data and MC. In the second case, the systematics from the large trigger counters' alignment are computed by varying the positions of S$_{4}$ and S$_\mu$ by $\mathcal{O}(\pm2\ \text{mm})$ along the axis of deflection and observe the change in the signal yield for different $Z'$ masses. The largest uncertainty is associated with masses $m_{Z'}<100$ MeV, since the trigger acceptance window corresponds to the tails of the distribution of the final-state muons momentum and angle \cite{Kirpichnikov:2021jev, Sieber:2023nkq}. This maximum value is conservatively chosen and found to be $5\%$.
\end{itemize}
To obtain the final systematic uncertainty on the signal yield $\mathcal{N}_\text{Z'}$, the contributions (i-iv) are added through quadrature, yielding a conservative total systematic of $8\%$. This estimate results in a $4\%$ uncertainty in the prediction of the bound on the coupling constant $g_{Z'}$.
\begin{table}[H]
    \centering
    \begin{tabular}{lr}
    \hline
    \hline
    Uncertainty source & Contribution \\
    \hline
    (i) Counting MOT & $\leq0.01$ \\
    (ii) $Z'$ physics & $\leq0.04$ \\
    (iii) Detectors' response & $\leq0.04$ \\
    (iv) Alignment effects & $\leq0.05$ \\
    \hline
    Total (conservatively) & $\leq0.08$ \\
    \hline
    \hline
    \end{tabular}
    \caption{\label{table:uncertainties}Main sources of uncertainty contributing to the systematics on the signal yield. The uncertainties are added through quadrature.}
\end{table}
\section{Results\label{sec:results}}
The sensitivity to New Physics of the NA64 experiment running in muon mode is based on the computation of the upper limits on, among other scenarios, the $Z'$ production. Those are computed at 90\% confidence levels (C.L.) following the modified frequentist approach \cite{Lista:2016tva}, using the \texttt{RooFit}/\texttt{RooStats} \cite{Verkerke:2003ir,Wolffs:2022fkh, Moneta:2010pm} profile likelihood ratio statistical test in the asymptotic approximation \cite{Cowan:2010js}. This procedure is encompassed in the CMS \cite{CMS:2008xjf} \texttt{combine} \cite{Hayrapetyan:2895097} analysis software. The total number of signal events falling within the signal box is estimated from Eq. \eqref{eq:signal-yield} from the sum of the two trigger configurations (one for each bin), such that \cite{Andreev:2024sgn}
\begin{equation}
    \mathcal{N}_{Z'}=\sum_{t=1,2}\mathcal{N}_{Z'}^{t}=\sum_{t=1,2}N_\text{MOT}^{t}\times\kappa_\text{tot}^{t}\times n_{Z^\prime}^{t}(m_{Z'},g_{Z'}),
\end{equation}
with $N_\text{MOT}^{t}$ the number of MOT for trigger configuration $t$, $\kappa_\text{tot}^{t}=\kappa_{\text{S}_{1}\text{S}_{0}}\kappa_{Z'}^{t}$ the total signal efficiency and $n_{Z^\prime}^{t}$ the mass- and coupling-dependent number of $Z'$ produced in the ECAL target per MOT. These values are extracted from both data and MC as described in Secs. \ref{sec:mc-approach}, \ref{sec:data-analysis} and \ref{sec:signal-yield}, to properly reproduce the two trigger configurations running conditions. Additionally, both the background estimate and its uncertainty, as well as the signal systematics described in Secs. \ref{sec:background-sources} and \ref{sec:systematics}, are taken into account when computing the upper limits.
\subsection{Constraints on the muon $(g-2)_\mu$}
The upper limits for the $Z'$ vector boson from the vanilla $L_\mu-L_\tau$ model are computed in the parameter space compatible with the muon $(g-2)_\mu$ anomaly. The purely invisible decay channel $Z'\rightarrow\bar{\nu}\nu$ is assumed, considering the proper branching ratio for visibly decaying events, $Z'\rightarrow\mu^{+}\mu^{-}$, above $2m_\mu$ (see Eq. \eqref{eq:decays-neutrinos}). The $g-2$ band is computed based on Eq. \eqref{eq:alpha-light-vector}, within $2\sigma$, using as reference value for $a_\mu(\text{Exp})=116\ 592\ 059(22)\times10^{-11}$ the latest results from the Muon $g-2$ collaboration at Fermilab \cite{Muong-2:2023cdq}, and the Muon $g-2$ Theory Initiative recommended value for $a_\mu(\text{SM})=116\ 591\ 810(43)\times10^{-11}$ \cite{Aoyama:2020ynm}. \\ \indent
After unblinding, no event compatible with $Z'\rightarrow\text{invisible}$ is observed within the signal box (see Fig. \ref{fig:g-2-vector-scalar}, left), setting the most stringent limits on the $Z'$ as an explanation to the muon $(g-2)_\mu$ in the parameter space $m_{Z'}\gtrsim0.01$ GeV and $m_{Z'}\leq2m_\mu$ with $g_{Z'}\geq5\times10^{-4}$ \cite{Andreev:2024sgn}. As a comparison, constraints from previous experiments such as neutrino trident searches $\nu N\rightarrow\nu N\mu^{+}\mu^{-}$ with the CCFR experiment \cite{Altmannshofer:2014pba,CCFR:1991lpl} are shown. Additional limits with a $Z'$ mediator behaving as an intermediate virtual vector boson in neutrinos scattering off electrons, $\nu_\mu-e$ or $\nu_\tau-e$, are plotted for the BOREXINO detector \cite{Kamada:2015era,Kaneta:2016uyt,Gninenko:2020xys}, setting constraints on masses smaller than a few MeVs. In the higher mass range, the BaBar experiment sets limits \cite{Capdevilla:2021kcf} on the visible searches for $Z^\prime\rightarrow\mu^{+}\mu^{-}$ produced in electron-positron annihilation, $e^{+}e^{-}\rightarrow Z'\mu^{+}\mu^{-}$. The constraints from the latest Belle II results \cite{Belle-II:2019qfb} are also shown for direct electron-positron annihilation. For completeness, the recent limits on the vanilla $Z'$ scenario searches with the electron program of the NA64 experiment, $e^{-}N\rightarrow e^{-}NZ'$, are shown \cite{NA64:2022rme}.\\ \indent
\begin{widetext}
\par\smallskip\noindent
\centerline{\begin{minipage}{\linewidth}
\begin{figure}[H]
    \centering
    \includegraphics[width=0.45\textwidth]{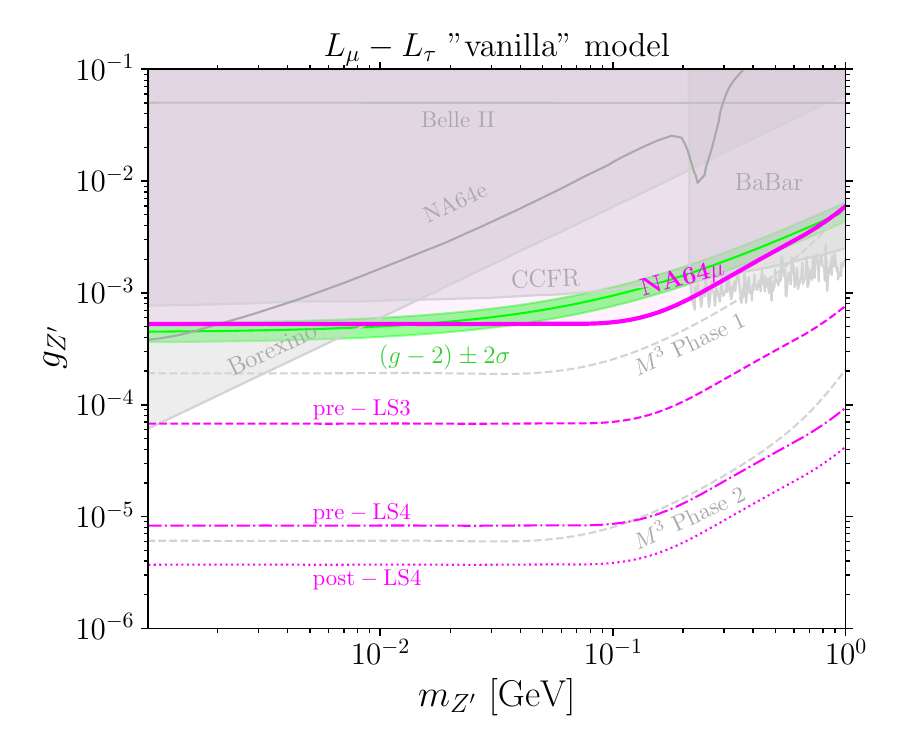}
    \hspace{6mm}
    \includegraphics[width=0.45\textwidth]{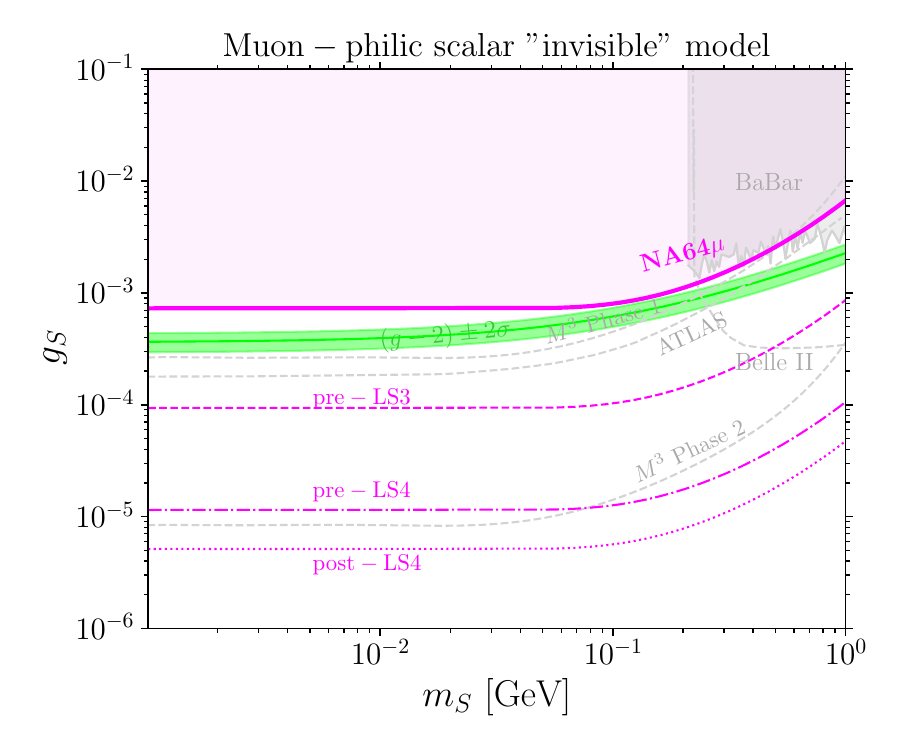}
    \caption{The NA64$\mu$ 90\% C.L. exclusion limits in the parameter space compatible with a light boson as an explanation for the muon $(g-2)_\mu$. Left: The $Z'$ vector boson parameter space $(m_{Z'},\ g_{Z'})$ together with existing constraints from neutrino experiments such as BOREXINO \cite{Kamada:2015era,Kaneta:2016uyt,Gninenko:2020xys} and CCFR \cite{Altmannshofer:2014pba,CCFR:1991lpl}, visible searches in electron-positron annihilation with BaBar \cite{Capdevilla:2021kcf}, Belle II constraints \cite{Belle-II:2019qfb} and the NA64 electron program limits \cite{Andreev:2024lps}. Projections for the pre-LS3, pre-LS4, and post-LS4 phases of the muon program are shown together with the $M^{3}$ missing momentum searches \cite{Kahn:2018cqs}. Right: The $S$ scalar boson parameter space $(m_{S},\ g_{S})$ together with existing constraints from BaBar and projections for the pre-LS3, pre-LS4, and post-LS4 phases of the muon program, as well as ATLAS HL-LHC \cite{Galon:2019owl} and $M^{3}$.}
    \label{fig:g-2-vector-scalar}
\end{figure}
\end{minipage}}
\end{widetext}
In addition, the projected sensitivities of the NA64 muon program are shown for both the LHC Run 3 (prior to the CERN Long Shutdown 3 (LS3)) and LHC Run 4 (after LS3). For completeness, the expected limits for the post-LS4 phase, LHC Run 5, are also computed. Those computations are based on the foreseen detectors' upgrade to cope with (i) higher beam intensity up to $10^{8}\ \mu$/spill to optimally exploit the M2 beam-line capabilities, as well as (ii) reducing the background levels discussed in Sec. \ref{sec:background-sources}. In the former case, this is achieved through an upgrade of the trigger system. In the second case, the momentum mis-reconstruction is reduced by the addition of a third magnet spectrometer upstream of the interaction point within the ECAL, effectively reducing the magnet lever-arm to $\simeq6$ m with respect to MS1 (BEND6) to better reconstruct $p_\text{in}$. The associated reduction of background can be extracted from the preliminary analysis of the 2023 data \cite{MolinaBueno:2868332} with a factor $<10^{-2}$. This additional magnet spectrometer also reduces the background from in-flight kaons decays given a shorter distance to the target by a factor $\sim10^{-1}$. It is worth noting that this background could be further reduced and controlled by proper identification of kaons along the beam-line through the use of Cherenkov counter with achromatic ring focus (CEDAR) \cite{NA62:2023mud}. Finally, the computations of the projected limits assume a reduction of the background related to non-hermeticity of the detectors'. The associated factor is estimated through a detailed MC study of the addition of a second prototype VHCAL to further reduce large-angle scattered secondaries from upstream interactions and is found to be $\leq10^{-1}$. The sensitivities are thus computed and shown in Fig. \ref{fig:g-2-vector-scalar} assuming a total statistics of respectively $3\times10^{11}$, $2\times10^{13}$ and $10^{14}$ MOT for the periods corresponding to LHC Runs 3, 4 and 5 (referred to pre-LS3, pre-LS4 and post-LS4 in this work), and a gain in efficiency $\simeq4$ as extracted from data \cite{MolinaBueno:2868332}. In the latter case, the accumulated statistics stands just below the irreducible background associated with the neutrino fog due to quasi-elastic charged current (CCQE) scattering on nucleon, estimated from \cite{Kahn:2018cqs} to be $\simeq3\times10^{-15}$ per MOT. The projections for LHC Runs 3 and 4 co-exist with the estimated limits from the $M^{3}$ experiment at Fermilab (FNAL) \cite{Kahn:2018cqs} in the search for $Z'\rightarrow\text{invisible}$ through the bremsstrahlung-like reaction $\mu N\rightarrow\mu NZ'$, with both its phases 1 and 2 foreseen to accumulate respectively $10^{10}$ and $10^{13}$ MOT. With a statistics of $\leq10^{11}$, NA64$\mu$ is expected to fully cover the parameter space compatible with a light $Z'$ vector boson as an explanation for the muon $g-2$. For completeness, it is worth noting that projections in the search for a light $U(1)$ $L_\mu-L_\tau$ $Z'$ vector boson as an explanation for the muon $(g-2)_\mu$ have been made by the DUNE experiment \cite{Ballett:2019xoj,Altmannshofer:2019zhy} through neutrino-electron scattering, $\nu-e$, or direct production, $\nu_\mu\rightarrow\nu_\mu\mu^{+}\mu^{-}$.\\ \indent
Similar limits are obtained for the sensitivity of the NA64 muon program to a light muon-philic scalar as an explanation to the $(g-2)_\mu$. The simulated signal yield is estimated in the WW approximation of the bremsstrahlung-like production process $\mu N\rightarrow\mu NS$ \cite{Sieber:2023nkq}, using a similar simulation framework as the one described in Sec. \ref{sec:mc-approach}. Within this model, the one-loop order contribution is obtained through the Yukawa-like interaction such that \cite{Chen:2015vqy,Leveille:1977rc,Lindner:2016bgg}
\begin{equation}
    \Delta a_\mu^{S}=\frac{g_S^2}{8\pi^2}\int_{0}^{1}dx\ \frac{m_\mu^2(1-x)(1-x^2)}{m_\mu^2(1-x)^2+m_S^2x},
\end{equation}
with $g_S$ the coupling to SM muons, $m_S$ the mass of the scalar mediator. The 90\% C.L. upper limits are shown in Fig. \ref{fig:g-2-vector-scalar}, right, together with the preferred $(g-2)\pm2\sigma$ band. Because of the difference in the amplitudes squared of the underlying production cross-sections between a scalar and vector particle \cite{Kirpichnikov:2021jev,Sieber:2023nkq}, the NA64 experiment is not sensitive to the $(g-2)_\mu$ parameter space yet, as shown for the existing BaBar limits for $m_S$ below 1 GeV for visible $Z'$ events, but is expected to fully cover this parameter space by the pre-LS3 phase. Projections for the Belle II experiment in search of $4\mu$ events are shown \cite{Capdevilla:2021kcf}, together with the sensitivity of the $M^{3}$ experiment \cite{Kahn:2018cqs}. Further projections from searches for $S$ with the High-Luminosity Large Hadron Collider (HL-LHC) project are shown using the ATLAS calorimeters as a muon fixed-target experiment \cite{Galon:2019owl} for a luminosity $\mathcal{L}_\text{LHC}=3$ ab$^{-1}$. This complements the projected sensitivities of the FASER experiment in the process $\mu N\rightarrow\mu NS$ for $\mathcal{L}_\text{LHC}=250$ fb$^{-1}$ and $\mathcal{L}_\text{LHC}=3$ ab$^{-1}$ (FASER$\nu$ and FASER$\nu 2$) \cite{Ariga:2023fjg}.
\subsection{Constraints on light thermal Dark Matter}
As discussed in Sec. \ref{sec:introduction}, DS present predictive scenarios for light thermal DM (LTDM) below the weak scale, in the vicinity of the GeV scale. In the following, we present exclusion limits for invisibly decaying mediators with mass hierarchy $m_\text{MED}>2m_\chi$, in the parameter space spanned by the dimensionless interaction strength $y$ and $m_{\chi}$ as shown in Eq. \eqref{eq:dimensionless-y}.
\subsubsection{The charged $U(1)$ $L_\mu-L_\tau$ model}
The 90\% C.L. upper limits in the $m_\chi-y$ parameter space for an invisibly decaying $Z'$ vector boson are shown in Fig. \ref{fig:thermal-vector-scalar} for the choice of parameters $m_{Z'}/m_\chi=3$ and the dark couplings $g_\chi=5\times10^{-2}$ (left) and $g_\chi=1$ (right), close to the perturbative limit. In the former case, the choice of mass ratio $m_{Z'}/m_\chi$ is well-motivated to allow for on-shell decays of $Z'\rightarrow\text{DM}$, while staying away from the resonant enhancement of the annihilation rate in the Early Universe at $m_{Z'}\simeq2m_\chi$ \cite{Berlin:2018bsc}. The choice of $g_{\chi}=1$ is chosen to illustrate the weakest bound on the model \cite{Kahn:2018cqs}. For completeness, the favored parameters for the thermal targets of complex scalar, (pseudo-)Dirac, or Majorana LTDM scenarios relic abundance are shown. Those are extracted from the integration of the underlying Boltzmann equation containing the corresponding annihilation cross-sections \cite{Berlin:2018bsc}. Our results cover an additional portion of the thermal parameter space, complementing the CCFR experiment's bounds \cite{Altmannshofer:2014pba} below masses $m_\chi\leq0.3$ GeV, and further constraining a freeze-out parameter $y\lesssim6\times10^{-12}$ for the common choice of $g_\chi=5\times10^{-2}$. For completeness, the projections for the pre-LS3, pre-LS4, and post-LS4 phases are shown, together with the expected bounds from the $M^{3}$ experimental program's phases 1 and 2 \cite{Kahn:2018cqs}. For the choice of parameter $g_\chi=5\times10^{-2}$, the LTDM parameter space is expected to be fully probed through the pre-LS4 period.
\begin{widetext}
\par\smallskip\noindent
\centerline{\begin{minipage}{\linewidth}
\begin{figure}[H]
    \centering
    \includegraphics[width=0.45\textwidth]{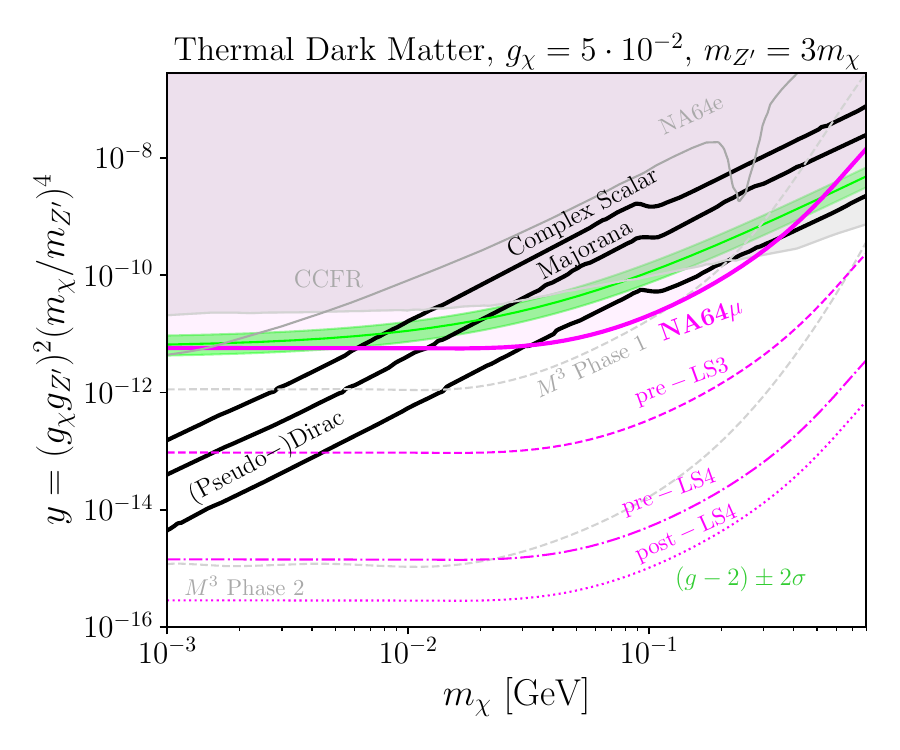}
    \hspace{6mm}
    \includegraphics[width=0.45\textwidth]{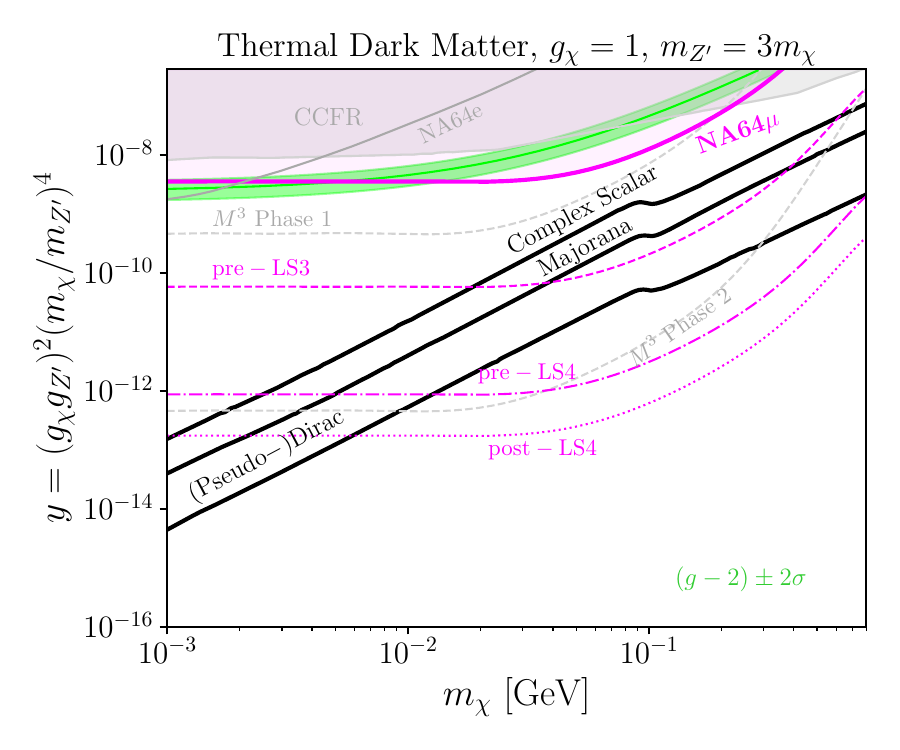}
    \caption{The NA64$\mu$ 90\% C.L. exclusion limits in the LTDM parameter space $y-m_{\chi}$ compatible with an invisibly decaying $Z'\rightarrow\text{DM}$ with (left) $g_\chi=5\times10^{-2}$ and (right) $g_\chi=1$ and mass ratio $m_\chi/m_{Z'}=3$. The existing constraints from the CCFR experiment \cite{Altmannshofer:2014pba,CCFR:1991lpl} are compared, and the thermal targets for complex scalar, (pseudo-)Dirac and Majorana thermal relics plotted \cite{Berlin:2018bsc}. Projections for the pre-LS3, pre-LS4, and post-LS4 periods of the muon program are shown together with the $M^{3}$ missing momentum searches \cite{Kahn:2018cqs}. The NA64 electron program limits are plotted for completeness \cite{Andreev:2024lps}.}
    \label{fig:thermal-vector-scalar}
\end{figure}
\end{minipage}}
\end{widetext}
\subsubsection{The dark photon $A'$ scenario}
\begin{figure}[H]
    \centering
    \includegraphics[width=0.44\textwidth]{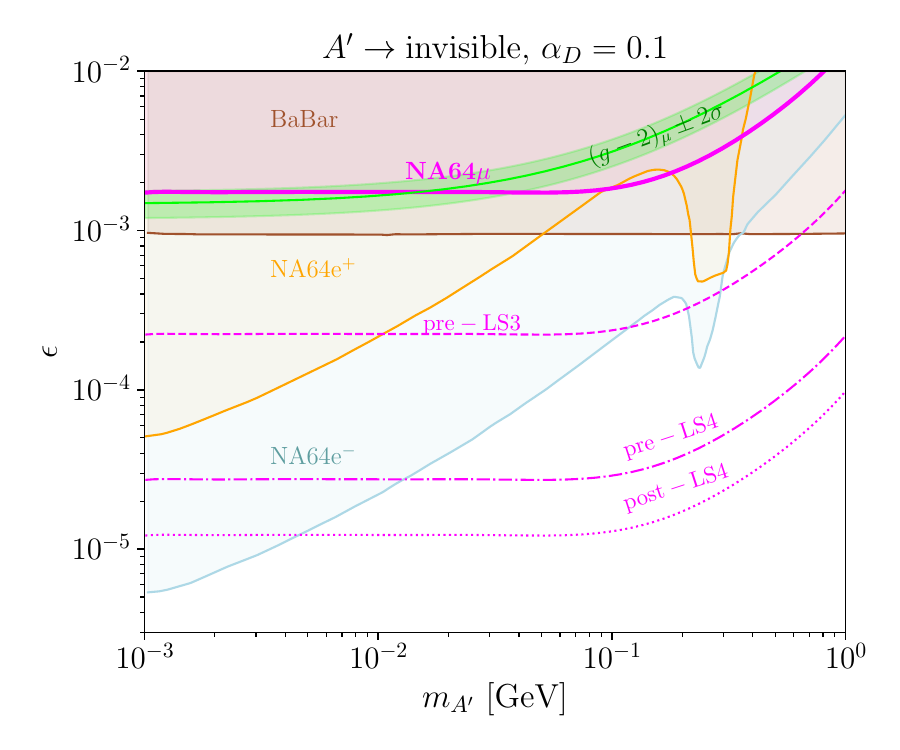}
    \caption{The NA64$\mu$ 90\% C.L. exclusion limits on the dark photon scenario, $A'\rightarrow\text{invisible}$. The $(m_{A'},\ \epsilon)$ parameter space is shown, together with the latest results from NA64$e^{-}$ \cite{NA64:2023wbi} and NA64$e^{+}$ \cite{NA64:2023ehh} and the existing limits from BaBar \cite{BaBar:2017tiz}. The peak is related to fermionic DM assuming $\alpha_D=0.1$. Projections for the pre-LS3, pre-LS4, and post-LS4 periods are shown.}
    \label{fig:dark-photon-g-2}
\end{figure}

The physics program of the NA64 experiment running in muon mode also aims at complementing \cite{Gninenko:2019qiv} the other beam dump modes with electrons \cite{NA64:2023wbi} and positions \cite{NA64:2023ehh}. Both modes study, among others, the bremsstrahlung-like production of the dark photon, $A'$, for which the mixing with photons, $\gamma-A'$, leads to a non-zero interaction with electrically charged SM particles, with charge $\epsilon e$, and interaction Lagrangian
\begin{equation}
    \mathcal{L}=\epsilon e A_\mu^{\prime}J_\text{EM}^\mu,
\end{equation}
with $e$ the electric charge, $A_\mu^\prime$ the massive vector field associated with the dark photon, and $J_\text{EM}^\mu$ the electromagnetic current. While $A'$ has been ruled out as an explanation for the muon $(g-2)_\mu$ (see e.g. \cite{NA64:2017vtt}), the NA64$\mu$ limits are illustratively shown in Fig. \ref{fig:dark-photon-g-2}, compared to the latest NA64 results with a total accumulated statistics of $9.37\times10^{11}$ electrons on target (EOT) and $1.01\times10^{10}$ positrons on target ($e^{+}$OT), for fermionic DM with $\alpha_D=0.1$, as well as existing constraints from BaBar \cite{BaBar:2017tiz}. Those are obtained at 90\% C.L. using the WW approximation \cite{Gninenko:2017yus} for the computations of the $A'$ signal yield, properly substituting $g_{Z'}\rightarrow g_{Z'}/e$. For completeness, the projections with $3\times10^{10}$, $2\times10^{13}$ and $10^{14}$ MOT are plotted to illustrate complementarity at high $m_{A'}$.\\ \indent
NA64$\mu$ limits in probing LTDM relic abundance in the scenario $A'\rightarrow\text{invisible}$ are shown in Fig. \ref{fig:aprime-thermal}, left and right, in the parameter space $y-m_\chi$, with $(g_\chi g_{Z'})^2\rightarrow\epsilon^2\alpha_D$ in Eq. \eqref{eq:dimensionless-y}, for the common choice of parameters $m_{A'}/m_\chi=3$ and $\alpha_D=0.1$, $\alpha_D=0.5$ respectively \cite{Battaglieri:2017aum,Beacham:2019nyx}. As shown from the projections, the complementarity to the NA64 $e^{-}$ and $e^{+}$ modes in probing the thermal targets is achieved with more statistics in the high $m_\chi$ masses region and fully probes the relic abundance through both the pre-LS3, pre-LS4, and post-LS4 phases for a choice of $\alpha_D=0.1$.
\begin{widetext}
\par\smallskip\noindent
\centerline{\begin{minipage}{\linewidth}
\begin{figure}[H]
    \centering
    \includegraphics[width=0.45\textwidth]{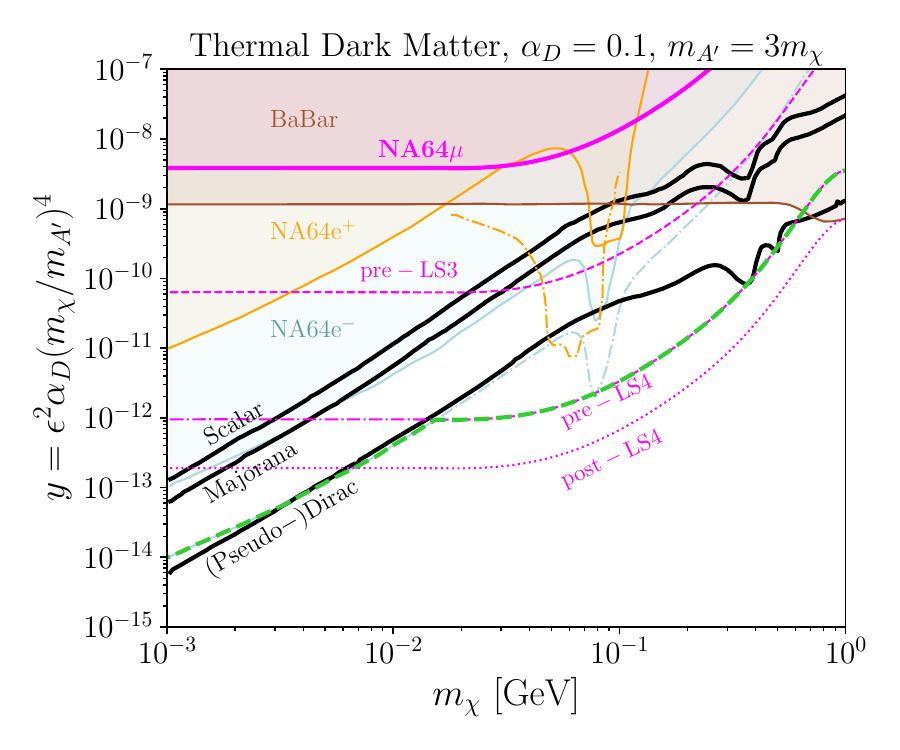}
    \hspace{6mm}
    \includegraphics[width=0.45\textwidth]{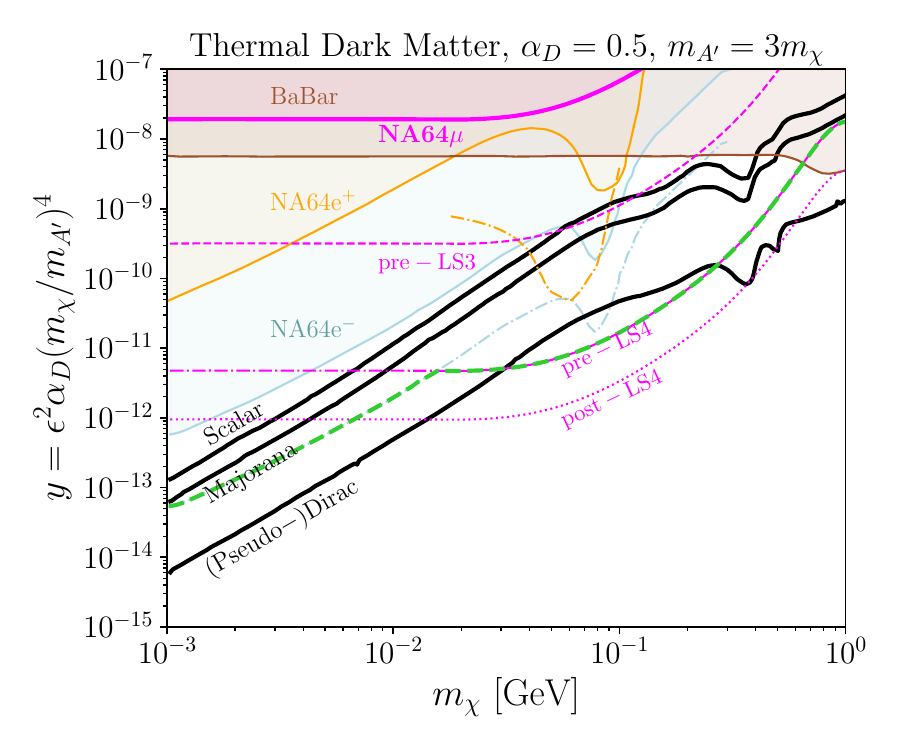}
    \caption{The NA64$\mu$ 90\% C.L. exclusion limits on the dark photon scenario, $A'\rightarrow\text{invisible}$ in the $(m_{\chi},\ y)$ parameter space, together with the DM target relic abundance for scalar, (pseudo-)Dirac and Majorana scenarios \cite{Berlin:2018bsc}. Left: Scenario with $\alpha_D=0.1$. Right: Scenario with $\alpha_D=0.5$. Projections for the pre-LS3, pre-LS4, and post-LS4 periods are shown for completeness. The combined projected limits (green dashed curve) for NA64$e^{-},e^{+},\mu$ are plotted, using the projections for $10^{13}$ EOT and $10^{11}$ $e^{+}$OT.}
    \label{fig:aprime-thermal}
\end{figure}
\end{minipage}}
\end{widetext}
\subsubsection{Light spin-0 mediators}
\begin{figure}[H]
    \centering
    \includegraphics[width=0.44\textwidth]{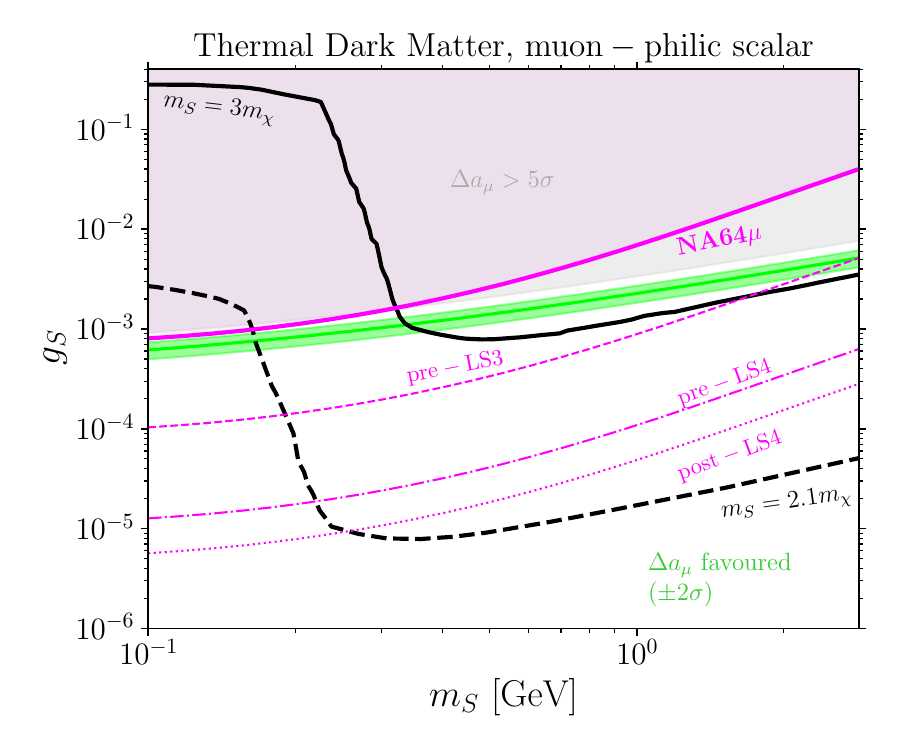}
    \caption{The NA64$\mu$ 90\% C.L. exclusion limits on invisibly decaying spin-0 scalar mediator, $S\rightarrow\text{invisible}$. The thermal targets for light DM are shown for respectively $m_S/m_\chi=3$ and $m_S/m_\chi=2.1$ with $g_\chi=1$, and extracted from \cite{Chen:2018vkr}. Projections for $3\times10^{11}$, $2\times10^{13}$ and $10^{14}$ MOT are plotted.}
    \label{fig:scalar-g-2-thermal}
\end{figure}

Constraints on lepto-philic DM are derived in the case of spin-0 muon-philic scalar mediator decaying invisibly, $S\rightarrow\text{invisible}$. The NA64$\mu$ 90\% C.L. upper limits on light DM are shown in Fig. \ref{fig:scalar-g-2-thermal} for the mass range of $m_S$ above a few hundred MeVs up to 3 GeV.\\ \indent
The choice of parameters is the dark coupling $g_\chi=1$ and the mass ratios $m_{S}/m_\chi=3$ and $m_{S}/m_\chi=2.1$ in the near-resonant regime for Dirac DM. The thermal targets are extracted from \cite{Chen:2018vkr}. With a total statistic of $1.98\times10^{10}$ MOT, the corresponding values of coupling $g_S\geq10^{-3}$ and mass $m_S\leq0.15$ GeV and $m_S\leq0.30$ GeV for respectively $m_S=2.1m_\chi$ and $m_S=3m_\chi$ are excluded.
\section{Conclusion\label{sec:conclusion}}
This work has presented in detail the first results in the search for DS through the missing energy/momentum technique at the CERN SPS. Having analyzed the full data sample of the 2022 run with a total statistics of $(1.98\pm0.02)\times10^{10}$ MOT, no evidence for the existence of a light $Z'\rightarrow\text{invisible}$ was found for the mass range $\leq1$ GeV. These results were obtained based on detailed computations of the single-differential and total production cross-sections in the Weisz\"{a}cker-Williams approximation \cite{Kirpichnikov:2021jev}, reproducing well the exact tree-level results up to a precision of $\leq 5\%$. Those were subsequently implemented in a computer-based program, \texttt{DMG4} \cite{Oberhauser:2024ozf}, to study both the signal event signature and yield within the whole \texttt{GEANT4}-based simulation of the experimental set-up. Based on previous MC-based studies \cite{Sieber:2021fue}, the trigger system was improved to compensate for the low production yield at high $Z'$ masses, $\sigma_{2\rightarrow3}^{Z'}\sim g_{Z'}^2\alpha^2 Z^2/m_{Z'}^{2}$, through the typical final-state muon emission angle, $\psi_\mu\sim m_{Z'}/E_\mu$. The simulations of both signal and SM events were validated with data, notably including a good agreement in the trigger rate $(0.026\pm0.004)\%$ that of calibration trigger, energy spectra in the ECAL and HCAL, and typical particle trajectory through the bending magnets. Having estimated both the total background level and signal systematics, constraints on the parameter space compatible with a light $Z'$ vector boson as an explanation for the muon $(g-2)_\mu$ were set through the 90\% C.L. upper limits on $g_{Z'}\geq5\times10^{-4}$ for masses $m_{Z'}\gtrsim0.01$ MeV up to $2m_\mu$, complementing existing constraints from BaBar and CCFR. Besides, new limits on the $(m_{S},\ g_S)$ parameter space associated with a scalar boson, $S$, were derived in the context of $(g-2)_\mu$.\\ \indent
The previous results of the NA64 muon program \cite{Andreev:2024sgn} on the parameter space associated with the dimensionless variable $y$, corresponding to the DM annihilation cross-section parameter, in the case where $Z^\prime\rightarrow \overline{\chi}\chi$, were extended to a scenario involving a light spin$-0$ scalar boson. For the choice of parameters $m_{S}=3m_\chi$, and close to the resonance, $m_{S}=2.1m_\chi$, those are found to constrain respectively masses $m_S\leq0.15$ GeV and $m_S\leq0.30$ GeV for the coupling $g_S\geq10^{-3}$.\\ \indent
To illustrate the complementarity of the NA64 muon program, the 90\% C.L. upper limits on the minimal dark photon scenario, $A'\rightarrow\text{invisible}$ were computed, and compared to the latest NA64 results for both the electron and positron modes. Although not covering new values of both parameter spaces $(m_{A'},\ \epsilon)$ and $(m_\chi,\ y)$ for the current statistics, the results indicate the possibility for additional coverage at high mass values due to the underlying cross-section behavior. In this regard, the projected sensitivities of NA64$\mu$ to the aforementioned DS scenarios are extracted for the estimated total statistics for the pre-LS3, pre-LS4, and post-LS4 periods with respectively $3\times10^{11}$, $2\times10^{13}$ and $10^{14}$ MOT. These results assume improvements in the experimental set-up, in particular with the trigger system to run at higher beam intensity, as well as with additional detectors such as a third magnet spectrometer in front of the ECAL target. With an increased coverage in the search for DM, NA64$\mu$ complement the worldwide effort for DS searches \cite{Jaeckel:2020dxj,Lanfranchi:2020crw,Krnjaic:2022ozp,Antel:2023hkf} through experimental program results such as those from Belle II \cite{Belle-II:2022yaw}, SHiP \cite{SHiP:2018yqc} and FASER \cite{Ariga:2023fjg}, and future projected sensitivities from $M^{3}$ \cite{Kahn:2018cqs}. It also paves the way to explore additional scenarios involving $\mu\rightarrow\tau$ or $\mu\rightarrow e$ lepton flavour conversion \cite{Gninenko:2018num,Gninenko:2022ttd,Radics:2023tkn} or millicharged particles \cite{Gninenko:2018ter}.
\section{Acknowledgments}
We gratefully acknowledge the support of the CERN management and staff, in particular the help of the CERN BE-EA department. We are grateful to C. Menezes Pires for his support with the beam momentum stations. We are also thankful for the contributions from HISKP, University of Bonn (Germany), ETH Zurich, and SNSF Grants No. 186181, No. 186158, No. 197346, No. 216602, (Switzerland), FONDECYT (Chile) under Grant
No. 1240066, ANID— Millennium Science Initiative Program—ICN2019 044 (Chile), RyC-030551-I, PID2021-123955NA-100 and CNS2022-135850 funded by MCIN/AEI/FEDER, UE (Spain). 
\bibliography{bibl}	

\end{document}